%**start of header
\documentstyle{amsppt}
\ifx\newheadisloaded\relax\immediate\write16{***already loaded} \else\let\newheadisloaded=\relax\fi
\gdef\islinuxolivetti{F}
\gdef\PSfonts{T}
%%%%%%%%%%%%%%%%%%%%%%%%%
%%%%layout %%%%%%%%%%%%%%
%%%%%%%%%%%%%%%%%%%%%%%%%
\magnification\magstep1
%for dvips with figures maybe uncomment next line

%%%%%%%%%%%%%%%%%%%%%%%
%%%% new parameters %%%
%%%%%%%%%%%%%%%%%%%%%%%
\newdimen\papwidth
\newdimen\papheight
\newskip\beforesectionskipamount  %how much to skip before section title
\newskip\sectionskipamount %how much to skip after section title
\def\sectionskip{\vskip\sectionskipamount}
\def\beforesectionskip{\vskip\beforesectionskipamount}
%%%%%%%%%%%%%%%%%%%%%%%
%%%% paper %%%%%%%%%%%%
%%%%%%%%%%%%%%%%%%%%%%%
\papwidth=16truecm
\if F\islinuxolivetti
\papheight=22truecm
\voffset=0.4truecm
\hoffset=0.4truecm
\else
\papheight=16truecm
\voffset=-1.5truecm
\hoffset=0.4truecm
\fi
%%%%%%%install variables%%%%%%%%%%%%%%%%%%%
\hsize=\papwidth
\vsize=\papheight
%%%%%%%%%%%%%%%%%%%%%%%%%
%%%%%% headline %%%%%%%%%
%%%%%%%%%%%%%%%%%%%%%%%%%
\catcode`\@=11
\ifx\amstexloaded@\relax
\else
\nopagenumbers
\headline={\ifnum\pageno>1 {\hss\tenrm-\ \folio\ -\hss} \else
{\hfill}\fi}
\fi
\catcode`\@=\active
%%%%%%%%%%%%%%%%%%%%%%%%%
\newdimen\texpscorrection
\texpscorrection=0.15truecm %must be 0.15truecm in ps_fonts
%%%%%%%%%%%%%%%%%%%%%%%%
%%%%%%% fontsizes %%%%%%
%%%%%%%%%%%%%%%%%%%%%%%%

\def\sectionsize{\twelvepoint}
\def\sectiontype{\bf}
\def\subsectionsize{}
\def\subsectiontype{\bf}
\def\em{\sl}%will be italic in reality
%%%%%%%%%%%%%%%%%%%%%
\newfam\truecmsy
\newfam\truecmr
\newfam\msbfam
\newfam\scriptfam
\newfam\truecmsy
%%%%%%%%%%%%%%%%%%%%%%%%%%%%%%%%%%%%%
\newskip\ttglue 
%%%%%%%%%%%%%%%%%%%%%%%%%%%%%%%%%%%%%%%%%%%
% Font for LINUX
%%%%%%%%%%%%%%%%%%%%%%%%%%%%%%%%%%%%%%%%%%%%%
\if T\islinuxolivetti
\papheight=11.5truecm
\fi
\if F\PSfonts
% Times-Roman
\font\twelverm=cmr12
\font\tenrm=cmr10
%\font\ninerm=cmr9
\font\eightrm=cmr8
\font\sevenrm=cmr7
\font\sixrm=cmr6
\font\fiverm=cmr5

% Times-Bold
\font\twelvebf=cmbx12
\font\tenbf=cmbx10
%\font\ninebf=cmbx9
\font\eightbf=cmbx8
\font\sevenbf=cmbx7
\font\sixbf=cmbx6
\font\fivebf=cmbx5

% Times-Italic
\font\twelveit=cmti12
\font\tenit=cmti10
%\font\nineit=cmti9
\font\eightit=cmti8
\font\sevenit=cmti7
\font\sixit=cmti6
\font\fiveit=cmti5

% Times-Oblique(slanted)
\font\twelvesl=cmsl12
\font\tensl=cmsl10
%\font\ninesl=cmsl9
\font\eightsl=cmsl8
\font\sevensl=cmsl7
\font\sixsl=cmsl6
\font\fivesl=cmsl5

% Math-Italic
\font\twelvei=cmmi12
\font\teni=cmmi10
%\font\ninei=cmmi9
\font\eighti=cmmi8
\font\seveni=cmmi7
\font\sixi=cmmi6
\font\fivei=cmmi5

% Math-Symbols
\font\twelvesy=cmsy10	at	12pt
\font\tensy=cmsy10
%\font\ninesy=cmsy9
\font\eightsy=cmsy8
\font\sevensy=cmsy7
\font\sixsy=cmsy6
\font\fivesy=cmsy5
\font\twelvetruecmsy=cmsy10	at	12pt
\font\tentruecmsy=cmsy10
%\font\ninetruecmsy=cmsy9
\font\eighttruecmsy=cmsy8
\font\seventruecmsy=cmsy7
\font\sixtruecmsy=cmsy6
\font\fivetruecmsy=cmsy5

% CM-Roman
\font\twelvetruecmr=cmr12
\font\tentruecmr=cmr10
%\font\ninetruecmr=cmr9
\font\eighttruecmr=cmr8
\font\seventruecmr=cmr7
\font\sixtruecmr=cmr6
\font\fivetruecmr=cmr5

% Math-Boldfaces
\font\twelvebf=cmbx12
\font\tenbf=cmbx10
%\font\ninebf=cmbx9
\font\eightbf=cmbx8
\font\sevenbf=cmbx7
\font\sixbf=cmbx6
\font\fivebf=cmbx5

% Teletype
\font\twelvett=cmtt12
\font\tentt=cmtt10
%\font\ninett=cmtt9
\font\eighttt=cmtt8

% Big Math Symbols
\font\twelveex=cmex10	at	12pt
\font\tenex=cmex10
%\font\nineex=cmex9

% AMS Math Symbols
\font\twelvemsb=msbm10	at	12pt
\font\tenmsb=msbm10
%\font\ninemsb=msbm9
\font\eightmsb=msbm8
\font\sevenmsb=msbm7
\font\sixmsb=msbm6
\font\fivemsb=msbm5

%%% Fraktur
%%\newfam\frakfam
%%\font\twelvefrm=eufm10	at	12pt
%%\font\tenfrm=eufm10
%%%\font\ninefrm=eufm9
%%\font\eightfrm=eufm8
%%\font\sevenfrm=eufm7
%%\font\sixfrm=eufm6
%%\font\fivefrm=eufm5
%%
%%% Bold Fraktur
%%\newfam\frakbfam
%%\font\twelvefrb=eufb10 at 12pt
%%\font\tenfrb=eufb10
%%%\font\ninefrb=eufb9
%%\font\eightfrb=eufb8
%%\font\sevenfrb=eufb7
%%\font\sixfrb=eufb6
%%\font\fivefrb=eufb5
%%
% Script-Faces
\font\twelvescr=eusm10 at 12pt
\font\tenscr=eusm10
%\font\ninescr=eusm9
\font\eightscr=eusm8
\font\sevenscr=eusm7
\font\sixscr=eusm6
\font\fivescr=eusm5
\fi
\if T\PSfonts
%%%%%%%%%%%%%%%%%%%%%%%%%%%%%%%%%%%%%%
% Font mapping for postscript fonts.
%%%%%%%%%%%%%%%%%%%%%%%%%%%%%%%%%%%%%%
% Times-Roman
\font\twelverm=ptmr	at	12pt
\font\tenrm=ptmr	at	10pt
%\font\ninerm=ptmr	at	9pt
\font\eightrm=ptmr	at	8pt
\font\sevenrm=ptmr	at	7pt
\font\sixrm=ptmr	at	6pt
\font\fiverm=ptmr	at	5pt

% Times-Bold
\font\twelvebf=ptmb	at	12pt
\font\tenbf=ptmb	at	10pt
%\font\ninebf=ptmb	at	9pt
\font\eightbf=ptmb	at	8pt
\font\sevenbf=ptmb	at	7pt
\font\sixbf=ptmb	at	6pt
\font\fivebf=ptmb	at	5pt

% Times-Italic
\font\twelveit=ptmri	at	12pt
\font\tenit=ptmri	at	10pt
%\font\nineit=ptmri	at	9pt
\font\eightit=ptmri	at	8pt
\font\sevenit=ptmri	at	7pt
\font\sixit=ptmri	at	6pt
\font\fiveit=ptmri	at	5pt

% Times-Oblique(slanted)
\font\twelvesl=ptmro	at	12pt
\font\tensl=ptmro	at	10pt
%\font\ninesl=ptmro	at	9pt
\font\eightsl=ptmro	at	8pt
\font\sevensl=ptmro	at	7pt
\font\sixsl=ptmro	at	6pt
\font\fivesl=ptmro	at	5pt

% Math-Italic
\font\twelvei=cmmi12
\font\teni=cmmi10
%\font\ninei=cmmi9
\font\eighti=cmmi8
\font\seveni=cmmi7
\font\sixi=cmmi6
\font\fivei=cmmi5

% Math-Symbols
\font\twelvesy=cmsy10	at	12pt
\font\tensy=cmsy10
%\font\ninesy=cmsy9
\font\eightsy=cmsy8
\font\sevensy=cmsy7
\font\sixsy=cmsy6
\font\fivesy=cmsy5
\font\twelvetruecmsy=cmsy10	at	12pt
\font\tentruecmsy=cmsy10
%\font\ninetruecmsy=cmsy9
\font\eighttruecmsy=cmsy8
\font\seventruecmsy=cmsy7
\font\sixtruecmsy=cmsy6
\font\fivetruecmsy=cmsy5

% CM-Roman
\font\twelvetruecmr=cmr12
\font\tentruecmr=cmr10
%\font\ninetruecmr=cmr9
\font\eighttruecmr=cmr8
\font\seventruecmr=cmr7
\font\sixtruecmr=cmr6
\font\fivetruecmr=cmr5

% Math-Boldfaces
\font\twelvebf=cmbx12
\font\tenbf=cmbx10
%\font\ninebf=cmbx9
\font\eightbf=cmbx8
\font\sevenbf=cmbx7
\font\sixbf=cmbx6
\font\fivebf=cmbx5

% Teletype
\font\twelvett=cmtt12
\font\tentt=cmtt10
%\font\ninett=cmtt9
\font\eighttt=cmtt8

% Big Math Symbols
\font\twelveex=cmex10	at	12pt
\font\tenex=cmex10
%\font\nineex=cmex9

% AMS Math Symbols
\font\twelvemsb=msbm10	at	12pt
\font\tenmsb=msbm10
%\font\ninemsb=msbm9
\font\eightmsb=msbm8
\font\sevenmsb=msbm7
\font\sixmsb=msbm6
\font\fivemsb=msbm5

%%% Fraktur
%%\newfam\frakfam
%%\font\twelvefrm=eufm10	at	12pt
%%\font\tenfrm=eufm10
%%%\font\ninefrm=eufm9
%%\font\eightfrm=eufm8
%%\font\sevenfrm=eufm7
%%\font\sixfrm=eufm6
%%\font\fivefrm=eufm5
%%
%%% Bold Fraktur
%%\newfam\frakbfam
%%\font\twelvefrb=eufb10 at 12pt
%%\font\tenfrb=eufb10
%%%\font\ninefrb=eufb9
%%\font\eightfrb=eufb8
%%\font\sevenfrb=eufb7
%%\font\sixfrb=eufb6
%%\font\fivefrb=eufb5
%%
% Script-Faces
\font\twelvescr=eusm10 at 12pt
\font\tenscr=eusm10
%\font\ninescr=eusm9
\font\eightscr=eusm8
\font\sevenscr=eusm7
\font\sixscr=eusm6
\font\fivescr=eusm5
\fi
%%%%%%%%%%%%%%%%%%%%%%%%%
%%%%preloaded fonts%%%%%%
%%%%%%%%%%%%%%%%%%%%%%%%%
\def\eightpoint{\def\rm{\fam0\eightrm}%
\textfont0=\eightrm
  \scriptfont0=\sixrm
  \scriptscriptfont0=\fiverm 
\textfont1=\eighti
  \scriptfont1=\sixi
  \scriptscriptfont1=\fivei 
\textfont2=\eightsy
  \scriptfont2=\sixsy
  \scriptscriptfont2=\fivesy 
\textfont3=\tenex
  \scriptfont3=\tenex
  \scriptscriptfont3=\tenex 
\textfont\itfam=\eightit
  \scriptfont\itfam=\sixit
  \scriptscriptfont\itfam=\fiveit 
  \def\it{\fam\itfam\eightit}%
\textfont\slfam=\eightsl
  \scriptfont\slfam=\sixsl
  \scriptscriptfont\slfam=\fivesl 
  \def\sl{\fam\slfam\eightsl}%
\textfont\ttfam=\eighttt
  \def\tt{\fam\ttfam\eighttt}%
\textfont\bffam=\eightbf
  \scriptfont\bffam=\sixbf
  \scriptscriptfont\bffam=\fivebf
  \def\bf{\fam\bffam\eightbf}%
%%\textfont\frakfam=\eightfrm
%%  \scriptfont\frakfam=\sixfrm
%%  \scriptscriptfont\frakfam=\fivefrm
%%  \def\frak{\fam\frakfam\eightfrm}%
%%\textfont\frakbfam=\eightfrb
%%  \scriptfont\frakbfam=\sixfrb
%%  \scriptscriptfont\frakbfam=\fivefrb
%%  \def\bfrak{\fam\frakbfam\eightfrb}%
\textfont\scriptfam=\eightscr
  \scriptfont\scriptfam=\sixscr
  \scriptscriptfont\scriptfam=\fivescr
  \def\script{\fam\scriptfam\eightscr}%
\textfont\msbfam=\eightmsb
  \scriptfont\msbfam=\sixmsb
  \scriptscriptfont\msbfam=\fivemsb
  \def\bb{\fam\msbfam\eightmsb}%
\textfont\truecmr=\eighttruecmr
  \scriptfont\truecmr=\sixtruecmr
  \scriptscriptfont\truecmr=\fivetruecmr
  \def\truerm{\fam\truecmr\eighttruecmr}%
\textfont\truecmsy=\eighttruecmsy
  \scriptfont\truecmsy=\sixtruecmsy
  \scriptscriptfont\truecmsy=\fivetruecmsy
\tt \ttglue=.5em plus.25em minus.15em 
\normalbaselineskip=9pt
\setbox\strutbox=\hbox{\vrule height7pt depth2pt width0pt}%
\normalbaselines
\rm
}

\def\tenpoint{\def\rm{\fam0\tenrm}%
\textfont0=\tenrm
  \scriptfont0=\sevenrm
  \scriptscriptfont0=\fiverm 
\textfont1=\teni
  \scriptfont1=\seveni
  \scriptscriptfont1=\fivei 
\textfont2=\tensy
  \scriptfont2=\sevensy
  \scriptscriptfont2=\fivesy 
\textfont3=\tenex
  \scriptfont3=\tenex
  \scriptscriptfont3=\tenex 
\textfont\itfam=\tenit
  \scriptfont\itfam=\sevenit
  \scriptscriptfont\itfam=\fiveit 
  \def\it{\fam\itfam\tenit}%
\textfont\slfam=\tensl
  \scriptfont\slfam=\sevensl
  \scriptscriptfont\slfam=\fivesl 
  \def\sl{\fam\slfam\tensl}%
\textfont\ttfam=\tentt
  \def\tt{\fam\ttfam\tentt}%
\textfont\bffam=\tenbf
  \scriptfont\bffam=\sevenbf
  \scriptscriptfont\bffam=\fivebf
  \def\bf{\fam\bffam\tenbf}%
%%\textfont\frakfam=\tenfrm
%%  \scriptfont\frakfam=\sevenfrm
%%  \scriptscriptfont\frakfam=\fivefrm
%%  \def\frak{\fam\frakfam\tenfrm}%
%%\textfont\frakbfam=\tenfrb
%%  \scriptfont\frakbfam=\sevenfrb
%%  \scriptscriptfont\frakbfam=\fivefrb
%%  \def\bfrak{\fam\frakbfam\tenfrb}%
\textfont\scriptfam=\tenscr
  \scriptfont\scriptfam=\sevenscr
  \scriptscriptfont\scriptfam=\fivescr
  \def\script{\fam\scriptfam\tenscr}%
\textfont\msbfam=\tenmsb
  \scriptfont\msbfam=\sevenmsb
  \scriptscriptfont\msbfam=\fivemsb
  \def\bb{\fam\msbfam\tenmsb}%
\textfont\truecmr=\tentruecmr
  \scriptfont\truecmr=\seventruecmr
  \scriptscriptfont\truecmr=\fivetruecmr
  \def\truerm{\fam\truecmr\tentruecmr}%
\textfont\truecmsy=\tentruecmsy
  \scriptfont\truecmsy=\seventruecmsy
  \scriptscriptfont\truecmsy=\fivetruecmsy
\tt \ttglue=.5em plus.25em minus.15em 
\normalbaselineskip=12pt
\setbox\strutbox=\hbox{\vrule height8.5pt depth3.5pt width0pt}%
\normalbaselines
\rm
}

\def\twelvepoint{\def\rm{\fam0\twelverm}%
\textfont0=\twelverm
  \scriptfont0=\tenrm
  \scriptscriptfont0=\eightrm 
\textfont1=\twelvei
  \scriptfont1=\teni
  \scriptscriptfont1=\eighti 
\textfont2=\twelvesy
  \scriptfont2=\tensy
  \scriptscriptfont2=\eightsy 
\textfont3=\twelveex
  \scriptfont3=\twelveex
  \scriptscriptfont3=\twelveex 
\textfont\itfam=\twelveit
  \scriptfont\itfam=\tenit
  \scriptscriptfont\itfam=\eightit 
  \def\it{\fam\itfam\twelveit}%
\textfont\slfam=\twelvesl
  \scriptfont\slfam=\tensl
  \scriptscriptfont\slfam=\eightsl 
  \def\sl{\fam\slfam\twelvesl}%
\textfont\ttfam=\twelvett
  \def\tt{\fam\ttfam\twelvett}%
\textfont\bffam=\twelvebf
  \scriptfont\bffam=\tenbf
  \scriptscriptfont\bffam=\eightbf
  \def\bf{\fam\bffam\twelvebf}%
%%\textfont\frakfam=\twelvefrm
%%  \scriptfont\frakfam=\tenfrm
%%  \scriptscriptfont\frakfam=\eightfrm
%%  \def\frak{\fam\frakfam\twelvefrm}%
%%\textfont\frakbfam=\twelvefrb
%%  \scriptfont\frakbfam=\tenfrb
%%  \scriptscriptfont\frakbfam=\eightfrb
%%  \def\bfrak{\fam\frakbfam\twelvefrb}%
\textfont\scriptfam=\twelvescr
  \scriptfont\scriptfam=\tenscr
  \scriptscriptfont\scriptfam=\eightscr
  \def\script{\fam\scriptfam\twelvescr}%
\textfont\msbfam=\twelvemsb
  \scriptfont\msbfam=\tenmsb
  \scriptscriptfont\msbfam=\eightmsb
  \def\bb{\fam\msbfam\twelvemsb}%
\textfont\truecmr=\twelvetruecmr
  \scriptfont\truecmr=\tentruecmr
  \scriptscriptfont\truecmr=\eighttruecmr
  \def\truerm{\fam\truecmr\twelvetruecmr}%
\textfont\truecmsy=\twelvetruecmsy
  \scriptfont\truecmsy=\tentruecmsy
  \scriptscriptfont\truecmsy=\eighttruecmsy
\tt \ttglue=.5em plus.25em minus.15em 
\setbox\strutbox=\hbox{\vrule height7pt depth2pt width0pt}%
\normalbaselineskip=15pt
\normalbaselines
\rm
}
%
%%%%%constant subscript positions%%%%%
\fontdimen16\tensy=2.7pt
%\fontdimen13\tensy=2.7pt
\fontdimen13\tensy=4.3pt
\fontdimen17\tensy=2.7pt
\fontdimen14\tensy=4.3pt
\fontdimen18\tensy=4.3pt
\fontdimen16\eightsy=2.7pt
\fontdimen13\eightsy=4.3pt
\fontdimen17\eightsy=2.7pt
\fontdimen14\eightsy=4.3pt
\fontdimen18\sevensy=4.3pt
\fontdimen16\sevensy=1.8pt
\fontdimen13\sevensy=4.3pt
\fontdimen17\sevensy=2.7pt
\fontdimen14\sevensy=4.3pt
\fontdimen18\sevensy=4.3pt
%
%%%%%%%%%%%%%%%%%%%%%%%%%%%%%%%%%%%%%%%%%%%%%%%%%%%%%%%%%%%%%
%%%%%%%%%%%%%% redefine some math so that it is cmr %%%%%%%%%
%%%%%%%%%%%%%%%%%%%%%%%%%%%%%%%%%%%%%%%%%%%%%%%%%%%%%%%%%%%%%
\def\hexnumber#1{\ifcase#1 0\or1\or2\or3\or4\or5\or6\or7\or8\or9\or
 A\or B\or C\or D\or E\or F\fi}
\mathcode`\=="3\hexnumber\truecmr3D
\mathchardef\not="3\hexnumber\truecmsy36
\mathcode`\+="2\hexnumber\truecmr2B
\mathcode`\(="4\hexnumber\truecmr28
\mathcode`\)="5\hexnumber\truecmr29
\mathcode`\!="5\hexnumber\truecmr21
\mathcode`\(="4\hexnumber\truecmr28
\mathcode`\)="5\hexnumber\truecmr29
%\chardef`,="0\hexnum\truecmr3B

\def\Phi{\mathchar"0\hexnumber\truecmr08 }
\def\Gamma {\mathchar"0\hexnumber\truecmr00 }
\def\Delta {\mathchar"0\hexnumber\truecmr01 }
\def\Theta {\mathchar"0\hexnumber\truecmr02 }
\def\Lambda{\mathchar"0\hexnumber\truecmr03 }
\def\Xi {\mathchar"0\hexnumber\truecmr04 }
\def\Pi{\mathchar"0\hexnumber\truecmr05 }
\def\Sigma{\mathchar"0\hexnumber\truecmr06 }
\def\Upsilon {\mathchar"0\hexnumber\truecmr07 }
\def\Phi {\mathchar"0\hexnumber\truecmr08 }
\def\Psi {\mathchar"0\hexnumber\truecmr09 }
\def\Omega{\mathchar"0\hexnumber\truecmr0A }
%%%%%%%%%%%%%%%%%%%%%%%%%%%%%%%%%%%%%%%%%%%%%%
%%% macros  for cross reference %%%%%%%%%%%%%%
%%%%%%%%%%%%%%%%%%%%%%%%%%%%%%%%%%%%%%%%%%%%%%
%%
%%  counters %%%
%%  
\newcount\EQNcount \EQNcount=1
\newcount\CLAIMcount \CLAIMcount=1
\newcount\SECTIONcount \SECTIONcount=0
\newcount\SUBSECTIONcount \SUBSECTIONcount=1
%%
%% defining the symbolic value
%%
\def\ifff(#1,#2,#3){\ifundefined{#1#2}%
\expandafter\xdef\csname #1#2\endcsname{#3}\else%
\immediate\write16{!!!!!doubly defined #1,#2}\fi}
\def\NEWDEF #1,#2,#3 {\ifff({#1},{#2},{#3})}
\def\actualnumber{\number\SECTIONcount}
\def\EQ(#1){\lmargin(#1)\eqno\tageck(#1)}
\def\NR(#1){&\lmargin(#1)\tageck(#1)\cr}  %the same as &\tageck(xx)\cr in eqalignno
\def\tageck(#1){\lmargin(#1)({\rm \actualnumber}.\number\EQNcount)
 \NEWDEF e,#1,(\actualnumber.\number\EQNcount)
\global\advance\EQNcount by 1
%\immediate\write16{ EQ \equ(#1):#1  }
}
\def\SECT(#1)#2\par{\lmargin(#1)\SECTION#2\par%
\NEWDEF s,#1,{\actualnumber}
}
\def\SUBSECT(#1)#2\par{\lmargin(#1)
\SUBSECTION#2\par 
\NEWDEF s,#1,{\actualnumber.\number\SUBSECTIONcount}
}
%%%% the actual macro %%%%%%
\def\CLAIM #1(#2) #3\par{
\vskip.1in\medbreak\noindent
{\lmargin(#2)\bf #1\ \actualnumber.\number\CLAIMcount.} {\sl #3}\par
\NEWDEF c,#2,{#1\ \actualnumber.\number\CLAIMcount}
\global\advance\CLAIMcount by 1
\ifdim\lastskip<\medskipamount
\removelastskip\penalty55\medskip\fi}
\def\CLAIMNONR #1(#2) #3\par{
\vskip.1in\medbreak\noindent
{\lmargin(#2)\bf #1.} {\sl #3}\par
\NEWDEF c,#2,{#1}
\global\advance\CLAIMcount by 1
\ifdim\lastskip<\medskipamount
\removelastskip\penalty55\medskip\fi}
\def\SECTION#1\par{\vskip0pt plus.3\vsize\penalty-75
    \vskip0pt plus -.3\vsize
    \global\advance\SECTIONcount by 1
    \beforesectionskip\noindent
{\sectionsize\sectiontype \actualnumber.\ #1}
    \EQNcount=1
    \CLAIMcount=1
    \SUBSECTIONcount=1
    \nobreak\sectionskip\noindent}
\def\SECTIONNONR#1\par{\vskip0pt plus.3\vsize\penalty-75
    \vskip0pt plus -.3\vsize
    \global\advance\SECTIONcount by 1
    \beforesectionskip\noindent
{\sectionsize\sectiontype  #1}
     \EQNcount=1
     \CLAIMcount=1
     \SUBSECTIONcount=1
     \nobreak\sectionskip\noindent}
\def\SUBSECTION#1\par{\vskip0pt plus.2\vsize\penalty-75%
    \vskip0pt plus -.2\vsize%
    \beforesectionskip\noindent%
{\subsectionsize\subsectiontype \actualnumber.\number\SUBSECTIONcount.\ #1}
    \global\advance\SUBSECTIONcount by 1
    \nobreak\sectionskip\noindent}
\def\SUBSECTIONNONR#1\par{\vskip0pt plus.2\vsize\penalty-75
    \vskip0pt plus -.2\vsize
\beforesectionskip\noindent
{\subsectionsize\subsectiontype #1}
    \nobreak\sectionskip\noindent\noindent}
%%
%%  referring to something
%%
\def\ifundefined#1{\expandafter\ifx\csname#1\endcsname\relax}
\def\equ(#1){\ifundefined{e#1}$\spadesuit$#1\else\csname e#1\endcsname\fi}
\def\clm(#1){\ifundefined{c#1}$\spadesuit$#1\else\csname c#1\endcsname\fi}
\def\sec(#1){\ifundefined{s#1}$\spadesuit$#1
\else Section \csname s#1\endcsname\fi}
\def\fig(#1){\ifundefined{fig#1}$\spadesuit$#1\else\csname fig#1\endcsname\fi}
%%%%%%%%%%%%%TITLE PAGE%%%%%%%%%%%%%%%%%%%%
\let\endarg=\par
\def\finish{\def\endarg{\par\endgroup}}
\def\start{\endarg\begingroup}

 \def\beginFROM{\start\parskip=0pt\vskip\baselineskip
\def\finish{\def\endarg{\egroup\par\endgroup}}
  \vbox\bgroup\obeylines\eightpoint\em\finish}

\def\ABSTRACT#1\par{
\vskip 1in {\noindent\sectionsize\sectiontype Abstract.} #1 \par}

%%%%%%%%%%% The today mechanism %%%%%%%%%%%%%%%%%
\def\TODAY{\number\day~\ifcase\month\or January \or February \or March \or
April \or May \or June
\or July \or August \or September \or October \or November \or December \fi
\number\year\timecount=\number\time
\divide\timecount by 60
}
\newcount\timecount
\def\DRAFT{\def\lmargin(##1){\strut\vadjust{\kern-\strutdepth
\vtop to \strutdepth{
\baselineskip\strutdepth\vss\rlap{\kern-1.2 truecm\eightpoint{##1}}}}}
\font\footfont=cmti7
\footline={{\footfont \hfil File:\jobname, \TODAY,  \number\timecount h}}
}
%%%subitem an item in a vbox%%%%
\newbox\strutboxJPE
\setbox\strutboxJPE=\hbox{\strut}
\def\subitem#1#2\par{\vskip\baselineskip\vskip-\ht\strutboxJPE{\item{#1}#2}}
\gdef\strutdepth{\dp\strutbox}
\def\lmargin(#1){}

\def\hexnumber#1{\ifcase#1 0\or1\or2\or3\or4\or5\or6\or7\or8\or9\or
 A\or B\or C\or D\or E\or F\fi}
\textfont\msbfam=\tenmsb
\scriptfont\msbfam=\sevenmsb
\scriptscriptfont\msbfam=\fivemsb
\mathchardef\varkappa="0\hexnumber\msbfam7B%
%%%%%%%%%%%%%%%%%%%%%%%%%%%%%%%%%%%%%%%%%%%%%%%%%%%%%%%%
%%%%%%%%%%  Figures %%%%%%%%%%%%%%%%%%%%%%%%%%%%%%%%%%%%
%%%%%%%%%%%%%%%%%%%%%%%%%%%%%%%%%%%%%%%%%%%%%%%%%%%%%%%%
\newcount\FIGUREcount \FIGUREcount=0
\newdimen\figcenter
\def\definefigure#1{\global\advance\FIGUREcount by 1%
\NEWDEF fig,#1,{Figure\ \number\FIGUREcount}
\immediate\write16{  FIG \number\FIGUREcount : #1}}
%%%%%%%%%%%%%%%%%%%%%%%%%%%%%%%%%%%%%%%%%%%%%%%
%figure 1=psfile=NAME 2=height (in cm) 3=width (in cm) 4=caption  
%%%%%%%%%%%%%%%%%%%%%%%%%%%%%%%%%%%%%%%%%%%%%%%%%%%%%%%
\def\figure #1 #2 #3 #4\cr{\null%
\definefigure{#1}
{\goodbreak\figcenter=\hsize\relax
\advance\figcenter by -#3truecm
\divide\figcenter by 2
\midinsert\vskip #2truecm\noindent\hskip\figcenter
\includegraphics{#1}\vskip 0.8truecm\noindent \vbox{\eightpoint\noindent
{\bf\fig(#1)}: #4}\endinsert}}
%%%%%%%%%%%%%%%%%%%%%%%%%%%%%%%%%%%%%%%%%%%%%%%%%%%%%%%%%
%figurewithtex 1=psfile=NAME 2=texfile 3=height (in cm) 4=width
%(in cm) 5=caption
%%%%%%%%%%%%%%%%%%%%%%%%%%%%%%%%%%%%%%%%%%%%%%%%%%%%%%%%%
\def\figurewithtex #1 #2 #3 #4 #5\cr{\null%
\definefigure{#1}
{\goodbreak\figcenter=\hsize\relax
\advance\figcenter by -#4truecm
\divide\figcenter by 2
\midinsert\vskip #3truecm\noindent\hskip\figcenter
\includegraphics{#1}{\hskip\texpscorrection\input #2 }\vskip 0.8truecm\noindent \vbox{\eightpoint\noindent
{\bf\fig(#1)}: #5}\endinsert}}
%%%%%%%%%%%%%%%%%%%%%%%%%%%%%%%%%%%%%%%%%%%%%%%%%%%%%%%%%%%%%%%%%%
%figurewithtexplus 1=psfile=NAME 2=texfile 3=height (in cm) 4=width
%(in cm) 5=dist figure-caption 6=caption
%%%%%%%%%%%%%%%%%%%%%%%%%%%%%%%%%%%%%%%%%%%%%%%%%%%%%%%%%%%%%%%%%%
\def\figurewithtexplus #1 #2 #3 #4 #5 #6\cr{\null%
\definefigure{#1}
{\goodbreak\figcenter=\hsize\relax
\advance\figcenter by -#4truecm
%\advance\figcenter by -#4truecm
\divide\figcenter by 2
\midinsert\vskip #3truecm\noindent\hskip\figcenter
\includegraphics{#1}{\hskip\texpscorrection\input #2 }\vskip #5truecm\noindent \vbox{\eightpoint\noindent
{\bf\fig(#1)}: #6}\endinsert}}
%%%%%%%%%%%%%%%%%%%%%%%%%%%%%%%%%%%%%%%%%%%%%%%%%%%%%%%
\catcode`@=11
\def\footnote#1{\let\@sf\empty % parameter #2 (the text) is read later
  \ifhmode\edef\@sf{\spacefactor\the\spacefactor}\/\fi
  #1\@sf\vfootnote{#1}}
\def\vfootnote#1{\insert\footins\bgroup\eightpoint
  \interlinepenalty\interfootnotelinepenalty
  \splittopskip\ht\strutbox % top baseline for broken footnotes
  \splitmaxdepth\dp\strutbox \floatingpenalty\@MM
  \leftskip\z@skip \rightskip\z@skip \spaceskip\z@skip \xspaceskip\z@skip
  \textindent{#1}\footstrut\futurelet\next\fo@t}
\def\fo@t{\ifcat\bgroup\noexpand\next \let\next\f@@t
  \else\let\next\f@t\fi \next}
\def\f@@t{\bgroup\aftergroup\@foot\let\next}
\def\f@t#1{#1\@foot}
\def\@foot{\strut\egroup}
\def\footstrut{\vbox to\splittopskip{}}
\skip\footins=\bigskipamount % space added when footnote is present
\count\footins=1000 % footnote magnification factor (1 to 1)
\dimen\footins=8in % maximum footnotes per page
\catcode`@=12 % at signs are no longer letters
%%%%%%%%%%%%%%%%%%%%%%%
%%%  math symbols %%%%%
%%%%%%%%%%%%%%%%%%%%%%%

%%%%%%%%%%%%%%%%%other%%%%%%%%%%%%%%%%%%%%

\def\QED{\hfill\smallskip
         \line{$\hfill{\vcenter{\vbox{\hrule height 0.2pt
	\hbox{\vrule width 0.2pt height 1.8ex \kern 1.8ex
		\vrule width 0.2pt}
	\hrule height 0.2pt}}}$
               \ \ \ \ \ \ }
         \bigskip}
\def\real{{\bf R}}

\def\PROOF{\medskip\noindent{\bf Proof.\ }}
\def\REMARK{\medskip\noindent{\bf Remark.\ }}
\def\LIKEREMARK#1{\medskip\noindent{\bf #1.\ }}
%%%%%%%%%%%%%%%%%%%%%%%%%%%%%%%%%%%%%%%%%%%%%%%%%%%
%%%% paragraphs ...                        %%%%%%%%
%%%%%%%%%%%%%%%%%%%%%%%%%%%%%%%%%%%%%%%%%%%%%%%%%%%
\normalbaselineskip=5.25mm
\baselineskip=5.25mm
\parskip=10pt
\beforesectionskipamount=24pt plus8pt minus8pt
\sectionskipamount=3pt plus1pt minus1pt
%\beforesectionskipamount=42pt plus5pt minus2pt
%\sectionskipamount=1truecm
%\overfullrule=0pt
%\hfuzz=2pt
\def\em{\it}
%%%%%%%%%%%%%%%%%%%%%%%%%%%%%%%%%%%%%%%%%%%%%%%%
% choice of default layout %%%%%%%%%%%%%%%%%%%%%
\tenpoint
\null
%%%%%%%%%%%%%%%%%%%%%%%%%%%%%%%%%%%%%%%%%%%%%%%
%%%%%%% exit here if amstex %%%%%%%%%%%%%%%%%%%
\catcode`\@=11
\ifx\amstexloaded@\relax\catcode`\@=\active
 \fi
\catcode`\@=\active
%%%%%%%%%%%%%%%%BIBLIOGRAPHY%%%%%%%%%%%%%%%%%%%%
%%%%%%%%%%%%%%%%%%%%%%%%%%%%%%%%%%%%%%%%%%%%%%%%
\def\period{\unskip.\spacefactor3000 { }}
%
% ...invisible stuff
%
\newbox\noboxJPE
\newbox\byboxJPE
\newbox\paperboxJPE
\newbox\yrboxJPE
\newbox\jourboxJPE
\newbox\pagesboxJPE
\newbox\volboxJPE
\newbox\preprintboxJPE
\newbox\toappearboxJPE
\newbox\bookboxJPE
\newbox\bybookboxJPE
\newbox\publisherboxJPE
\newbox\inprintboxJPE
\def\refclearJPE{
   \setbox\noboxJPE=\null             \gdef\isnoJPE{F}
   \setbox\byboxJPE=\null             \gdef\isbyJPE{F}
   \setbox\paperboxJPE=\null          \gdef\ispaperJPE{F}
   \setbox\yrboxJPE=\null             \gdef\isyrJPE{F}
   \setbox\jourboxJPE=\null           \gdef\isjourJPE{F}
   \setbox\pagesboxJPE=\null          \gdef\ispagesJPE{F}
   \setbox\volboxJPE=\null            \gdef\isvolJPE{F}
   \setbox\preprintboxJPE=\null       \gdef\ispreprintJPE{F}
   \setbox\toappearboxJPE=\null       \gdef\istoappearJPE{F}
   \setbox\inprintboxJPE=\null        \gdef\isinprintJPE{F}
   \setbox\bookboxJPE=\null           \gdef\isbookJPE{F}  \gdef\isinbookJPE{F}
     
   \setbox\bybookboxJPE=\null         \gdef\isbybookJPE{F}
   \setbox\publisherboxJPE=\null      \gdef\ispublisherJPE{F}
}

\def\ref{\refclearJPE\bgroup}
\def\no   {\egroup\gdef\isnoJPE{T}\setbox\noboxJPE=\hbox\bgroup}
\def\by   {\egroup\gdef\isbyJPE{T}\setbox\byboxJPE=\hbox\bgroup}
\def\paper{\egroup\gdef\ispaperJPE{T}\setbox\paperboxJPE=\hbox\bgroup}
\def\yr{\egroup\gdef\isyrJPE{T}\setbox\yrboxJPE=\hbox\bgroup}
\def\jour{\egroup\gdef\isjourJPE{T}\setbox\jourboxJPE=\hbox\bgroup}
\def\pages{\egroup\gdef\ispagesJPE{T}\setbox\pagesboxJPE=\hbox\bgroup}
\def\vol{\egroup\gdef\isvolJPE{T}\setbox\volboxJPE=\hbox\bgroup\bf}
\def\preprint{\egroup\gdef
\ispreprintJPE{T}\setbox\preprintboxJPE=\hbox\bgroup}
\def\toappear{\egroup\gdef
\istoappearJPE{T}\setbox\toappearboxJPE=\hbox\bgroup}
\def\inprint{\egroup\gdef
\isinprintJPE{T}\setbox\inprintboxJPE=\hbox\bgroup}
\def\book{\egroup\gdef\isbookJPE{T}\setbox\bookboxJPE=\hbox\bgroup\em}
\def\publisher{\egroup\gdef
\ispublisherJPE{T}\setbox\publisherboxJPE=\hbox\bgroup}
\def\inbook{\egroup\gdef\isinbookJPE{T}\setbox\bookboxJPE=\hbox\bgroup\em}
\def\bybook{\egroup\gdef\isbybookJPE{T}\setbox\bybookboxJPE=\hbox\bgroup}
\newdimen\refindent
\refindent=5em
\def\endref{\egroup \sfcode`.=1000
 \if T\isnoJPE
% \setbox0=\hbox{[\unhbox\noboxJPE\unskip]\hss\unskip\enspace}%
%   \ifdim\refindent<\wd0\relax
%      \message{\string\refno: reference is wider than
%               you pretended when using \string\widestlabel.}%
%   \fi
 \hangindent\refindent\hangafter=1
      \noindent\hbox to\refindent{[\unhbox\noboxJPE\unskip]\hss}\ignorespaces
     \else  \noindent    \fi
% \if T\isnoJPE  \item{[\unhbox\noboxJPE\unskip]}
%     \else  \noindent    \fi
 \if T\isbyJPE    \unhbox\byboxJPE\unskip: \fi
 \if T\ispaperJPE \unhbox\paperboxJPE\unskip\period \fi
 \if T\isbookJPE {\it\unhbox\bookboxJPE\unskip}\if T\ispublisherJPE, \else.
\fi\fi
 \if T\isinbookJPE In {\it\unhbox\bookboxJPE\unskip}\if T\isbybookJPE,
\else\period \fi\fi
 \if T\isbybookJPE  (\unhbox\bybookboxJPE\unskip)\period \fi
 \if T\ispublisherJPE \unhbox\publisherboxJPE\unskip \if T\isjourJPE, \else\if
T\isyrJPE \  \else\period \fi\fi\fi
 \if T\istoappearJPE (To appear)\period \fi
 \if T\ispreprintJPE Pre\-print\period \fi
 \if T\isjourJPE    \unhbox\jourboxJPE\unskip\ \fi
 \if T\isvolJPE     \unhbox\volboxJPE\unskip\if T\ispagesJPE, \else\ \fi\fi
 \if T\ispagesJPE   \unhbox\pagesboxJPE\unskip\  \fi
 \if T\isyrJPE      (\unhbox\yrboxJPE\unskip)\period \fi
 \if T\isinprintJPE (in print)\period \fi
\filbreak
}
%%%%%%%%%%%%%%%%%%%%%%%%%%%%%%%%%%%%%%%%%%%%%%%%%%%%
%%%%%EOF

\input xdefs
\NoBlackBoxes
\document
\TagsAsText
\TagsOnRight
%%%%%%%%%%%%%%%%%%%%%%%%%%%%%%%%%%%%%%%%%%%%%%%%%%

\undefine\diam
\define\diam{\operatorname{diam}}
\define\spt{\operatorname{spt}}
\define\por{\operatorname{por}}
\define\dimp{\operatorname{dim_p}}
\define\dimH{\operatorname{dim_H}}
\define\m(#1,#2){\mu\bigl (B(#1,#2)\bigr )}
\undefine\real
\define\real{\Bbb R}
\def\L{{\text{L}}}
\def\R{{\text{R}}}

%%%%%%%%%%%%%%%%%%%%%%%%%%%%%%%
\let\truett=\tt
\fontdimen3\tentt=2pt\fontdimen4\tentt=2pt
\def\tt{\hfill\break\null\kern -2truecm\truett **** }
\let\epsilon=\varepsilon
\undefine\EQ
\define\EQ(#1){\lmargin(#1)\tag\TAG(#1)}
\undefine\NR
\define\NR(#1){\lmargin(#1)\tag\TAG(#1)\cr}  %the same as &\TAG(xx)\cr in eqalignno
\define\TAG(#1){{{\text {\actualnumber}}.\number\EQNcount}%
\NEWDEF e,#1,(\actualnumber.\number\EQNcount)
\global\advance\EQNcount by 1%
}
%%%--\def\SECT(#1)#2\par{\lmargin(#1)\SECTION#2\par
%%%--\NEWDEF s,#1,{\actualnumber}\noindent
%%%--}
%%%--\def\SUBSECT(#1)#2\par{\lmargin(#1)
%%%--\SUBSECTION#2\par 
%%%--\NEWDEF s,#1,{\actualnumber.\number\SUBSECTIONcount}
%%%--}
%%%--%%%% the actual macro %%%%%%
\undefine\CLAIM
\def\CLAIM #1(#2) #3\par{
{\csname proclaim\endcsname{\bf #1
\actualnumber.\number\CLAIMcount}\lmargin(#2)#3\csname endproclaim\endcsname}
\NEWDEF c,#2,{#1\ \actualnumber.\number\CLAIMcount}
\global\advance\CLAIMcount by 1
}
\undefine\lmargin
\define\lmargin(#1){} %Put this as a comment if you use DRAFT
\undefine\DRAFT
\define\DRAFT{\define\lmargin(##1){\strut\vadjust{\kern-\strutdepth
\vtop to \strutdepth{
\baselineskip\strutdepth\vss\rlap{\kern-1.2 truecm\eightpoint{##1}}}}}
\font\footfont=cmti7
\footline={{\footfont \hfil File:\jobname, \TODAY,  \number\timecount h}}
}
%%%--\def\CLAIMNONR #1(#2) #3\par{
%%%--\vskip.1in\medbreak\noindent
%%%--{\lmargin(#2)\bf #1.} {\sl #3}\par
%%%--\NEWDEF c,#2,{#1}
%%%--\global\advance\CLAIMcount by 1
%%%--\ifdim\lastskip<\medskipamount
%%%--\removelastskip\penalty55\medskip\fi}
\def\SECTION#1\par{{
\global\advance\SECTIONcount by 1
\global\EQNcount=1
\global\CLAIMcount=1
\global\SUBSECTIONcount=1
\csname head\endcsname 
\actualnumber.\ #1 
\endhead}}
%%%--\def\SECTIONNONR#1\par{\vskip0pt plus.3\vsize\penalty-75
%%%--	 \vskip0pt plus -.3\vsize
%%%--	 \global\advance\SECTIONcount by 1
%%%--	 \beforesectionskip\noindent
%%%--{\sectionsize\sectiontype  #1}
%%%--	     \EQNcount=1
%%%--	     \CLAIMcount=1 
%%%--	     \SUBSECTIONcount=1
%%%--	     \nobreak\sectionskip\noindent}
%%%--\def\SUBSECTION#1\par{\vskip0pt plus.2\vsize\penalty-75%
%%%--	 \vskip0pt plus -.2\vsize%
%%%--	 \beforesectionskip\noindent%
%%%--{\subsectionsize\subsectiontype \actualnumber.\number\SUBSECTIONcount.\ #1}
%%%--	 \global\advance\SUBSECTIONcount by 1
%%%--	 \nobreak\sectionskip\noindent}
%%%--\def\SUBSECTIONNONR#1\par{\vskip0pt plus.2\vsize\penalty-75
%%%--	 \vskip0pt plus -.2\vsize
%%%--\beforesectionskip\noindent
%%%--{\subsectionsize\subsectiontype #1}
%%%--	 \nobreak\sectionskip\noindent\noindent}
%%%--%%
\def\LIKEREMARK#1{\remark{#1}}
\def\REMARK{\remark{Remark}}
%\normalbaselineskip=12pt
%\baselineskip=12pt
\parskip=0pt
\parindent=22.222pt

%%\\begin{document}%%% this is needed to trick emacs into finding header
%**end of header
%\DRAFT

\topmatter

\title Porosities and dimensions of measures
\endtitle
\author Jean-Pierre Eckmann$^1$, Esa J\"arvenp\"a\"a$^2$, and 
        Maarit J\"arvenp\"a\"a$^2$\endauthor
\affil University of Geneva, Departments of Physics and Mathematics,\\
  1211 Geneva 4, Switzerland$^1$\\
  University of Jyv\"askyl\"a, Department of Mathematics, P.O. Box 35,
  FIN-40351 Jyv\"askyl\"a, Finland$^2$\\ 
  Jean-Pierre.Eckmann\@physics.unige.ch, esaj\@math.jyu.fi, and 
  amj\@math.jyu.fi
  \endaffil

\leftheadtext\nofrills{J.-P. Eckmann, E. and M. J\"arvenp\"a\"a}
\rightheadtext\nofrills{Porosities and dimensions of measures}

\abstract We introduce a concept of porosity for measures and 
study relations between dimensions and porosities for two classes of
measures: measures on $\Bbb R^n$
which satisfy the doubling condition and strongly porous
measures on $\Bbb R$.
\endabstract
\endtopmatter  

\SECTION Introduction 

The aim of this paper is to relate porosity, as it can be
measured, to dimension. The requirement of obtaining information about
experimentally measurable objects leads us to consider measures, or 
mass distributions, rather than sets. 
For sets a relation between porosity and dimension
has been established by Mattila \cite{M1} and Salli \cite{S} using the
following definition of porosity:

\CLAIM Definition(defpor) The porosity of a set $A\subset\Bbb R^n$ at
a point $x\in\Bbb R^n$ is defined by
$$\por(A,x)\,=\,\liminf_{r\downarrow 0}\por(A,x,r)~,$$
where
$$\por(A,x,r)\,=\,\sup\{p\ge 0~:~\text{ there is }z\in\Bbb R^n
  \text{ such that }B(z,pr)\subset B(x,r)\setminus A\}~.$$
Here $B(x,r)$ is the closed ball with radius $r$ and with centre at $x$.
The porosity of $A\subset\Bbb R^n$ is
$$
\por(A)=\inf\{\por(A,x): x\in A\}~.
$$

\medskip

Clearly $0\le\por(A,x,r)\le\frac 12$ for $x\in A$, and so 
$0\le\por(A)\le\frac 12$.
The quantity $\por(A,x,r)$ gives the relative radius of the largest disk
which fits into $B(x,r)$ and which does not intersect $A$.
In this sense it gives the size of the biggest hole in $A$. 
 
For Hausdorff dimension, $\dimH$, 
it is not difficult to see that there exists a function 
$d:(0,\frac 12]\to (0,1]$ such that $\dimH(A)\le n-d(\por(A))$ for all
$A\subset\Bbb R^n$ (see \cite{S, Introduction}). However, this bound obtained
using simple methods is very crude when the porosity is close to $\frac 12$.
The following theorem by Salli \cite{S} 
gives a better connection between dimensions and porosities for sets. 
For the definition of packing dimension, $\dimp$, see
\cite{M2, Chapter 5} or \cite{Fa, Chapter 2}.

\CLAIM Theorem(mainset) There is a non-decreasing function 
$\Delta_n:[0,\frac 12]\to[0,1]$ 
satisfying
$$
\lim_{p\uparrow{1\over 2}} \Delta_n(p)\,=\,1
$$
such that
$$\dimp(A)\,\le\, n-\Delta_n(\por(A))~\EQ(ineqset)$$
for all $A\subset\Bbb R^n$.

\medskip

According to \clm(mainset) the packing dimension of any set in
$\Bbb R^n$ with
porosity close to $\frac 12$ can be only a little bit bigger than $n-1$.
There is an explicit expression for the function
$\Delta_n$ in \cite{S}: 
$$
\Delta_n(p)\,=\,\max\{1-\frac{c_n}{\log(1/(1-2p))}~,0\}~
$$
where $c_n$ is a constant depending only on $n$. Salli also proved that 
this function gives the optimal convergence rate by constructing 
for all $\frac14<p<\frac12$ sets $A_p$ with $\por(A_p)\ge p$ and
$\dimp(A_p)\ge n-1+\frac{b_n}{\log(1/(1-2p))}$
for some constant $b_n<c_n$.
Salli's proof works for box-counting dimension as well (for the definition see
\cite{M2, Chapter 5} or \cite{Fa, Chapter 2}), but then one has to 
assume that $A\subset\Bbb R^n$
is uniformly porous in the following sense: there is $R>0$ such that
$$
\por(A,x,r)\ge p\text{ for all }x\in A\text{ and for all }0<r\le R~.
$$
In an earlier work by Mattila \cite{M1} the analogue of \clm(mainset) was
proved for Hausdorff dimension using different methods than those of
Salli's. 

In this paper we address the problem of studying analogues of \clm(mainset) 
for measures. After introducing porosities of measures (see \clm(defpormu)) 
we prove that in $\Bbb R^n$ an analogue to \clm(mainset) holds for
measures which satisfy the doubling condition (see \clm(doubling)).
We also consider the class of strongly porous measures (see
\clm(onemain)) in $\Bbb R$. This article is organized as follows.
In addition to the necessary notation and definitions we discuss some
basic properties of porosities and state our main theorem in Section 2.
The next section is dedicated to the proof of the main results. In
Section 4 we consider the role of the doubling condition and in 
the last section we study the situation in the real line. 

\SECTION Notation and main results

We define the quantities we are working with. 
We begin with the definitions of Hausdorff and packing dimensions for
measures in terms of local dimensions:

\CLAIM Definition(defdimmu) Let $\mu$ be a finite Borel measure on 
$\real^n$. The lower and upper local dimensions of
$\mu$ at a point $x\in\real^n$ are 
$$
\underline
d(\mu,x)\,=\,\liminf_{r\downarrow0}{\log\mu(B(x,r))\over\log r}~
$$
and
$$
\overline d(\mu,x)\,=\,\limsup_{r\downarrow0}{\log
\mu(B(x,r))\over\log r}~.
$$
If $\underline d(\mu,x)=\overline d(\mu,x)$, the common value is called
the local dimension of $\mu$ at $x$ and is denoted by $d(\mu,x)$.
The Hausdorff and packing dimensions
of $\mu$ are defined by
$$\dimH(\mu)\,=\,\sup\{s\ge0~:~
  \underline d(\mu,x)\ge s\text{ for }\mu
  \text{-almost all }x\in\Bbb R^n\}~\EQ(HD)$$
and
$$
\dimp(\mu)\,=\,\sup\{s\ge0~:~
  \overline d(\mu,x)\ge s\text{ for }\mu
  \text{-almost all }x\in\Bbb R^n\}~.\EQ(PD)
$$

\medskip

The local dimensions describe the power law behaviour of $\mu$-measure
of balls with small radius. For $\mu$-almost all points the
lower local dimension is at least $\dimH(\mu)$ and the upper one is at
least $\dimp(\mu)$. Clearly $\dimH(\mu)\le\dimp(\mu)$.

\REMARK
We will need the following equivalent definitions of Hausdorff and
packing dimensions of measures 
in terms of dimensions of sets (see \cite{Fa, Proposition 10.2}). In fact,
$$
\dimH(\mu)\,=\,\inf\{\dimH(A)~:~ A\text{ is a Borel set with
}\mu(A)>0\}~\EQ(HD3)
$$
and
$$
\dimp(\mu)\,=\,\inf\{\dimp(A)~:~ A\text{ is a Borel set with
}\mu(A)>0\}~.\EQ(PD3)
$$

\medskip

The porosity of a finite Borel measure $\mu$ on $\Bbb R^n$ is defined
using the following quantities: for $x\in\Bbb R^n$ and $r,\varepsilon>0$
set 
$$\align
  \por(\mu,x,r,\varepsilon)\,
  =\,\sup\{p\ge 0~:~ &\text{ there is }z\in\Bbb R^n
   \text{ such that } B(z,pr)\subset B(x,r)\\
  &\text{ and }\mu(B(z,pr))\le\varepsilon\mu(B(x,r))\}~.
 \endalign
$$

\CLAIM Definition(defpormu) Let $\mu$ be a finite Borel measure on
$\Bbb R^n$. The porosity of $\mu$ at a point $x\in\Bbb R^n$
is defined by
$$
\por(\mu,x)\,=\,\lim_{\epsilon\downarrow 0}\liminf_{r\downarrow 0}
  \por(\mu,x,r,\varepsilon)~.
\EQ(porlocal)
$$
The porosity of $\mu$ is 
$$
\por(\mu)\,=\,\inf\{s\ge0: \por(\mu,x)\le s\text{ for }\mu\text{-almost all }
x\in\Bbb R^n\}~.\EQ(porglob)
$$

\medskip

In \equ(porlocal) the limit as $\varepsilon\downarrow0$ exists since
$\liminf_{r\downarrow 0}\por(\mu,x,r,\varepsilon)$ is non-decreasing
and bounded.

\REMARK 
1. We show now that the porosity of a measure has the same upper bound 
than that of a set, that is, $\por(\mu)\le\frac12$ for all finite Borel
measures $\mu$ on $\Bbb R^n$.
By \cite{C, (1.10)} for $\mu$-almost all $x\in\Bbb R^n$ we have
$\overline d(\mu,x)\le n$ giving 
$$\mu(B(x,r))\ge r^{2n}\EQ(lower)$$ 
for all sufficiently small $r>0$. Assume that there is such a point $x$ with 
$\por(\mu,x)>\frac 12(1+\delta)>\frac12$ for some 
$0<\delta<1$. Let $\varepsilon\le\delta^{3n}$ be sufficiently small. Then
for all sufficiently small $r>0$ there is $z\in\Bbb R^n$ such that 
$B(z,\frac r2(1+\delta))\subset B(x,r)$ and   
$\mu(B(z,\frac r2(1+\delta)))\le\varepsilon\mu(B(x,r))$. Hence for all
such $r$ we have $\mu(B(x,\delta r))\le\varepsilon\mu(B(x,r))$.
Iterating this $k$ times we obtain for all positive integers $k$
$$\mu(B(x,\delta^k r))\le\varepsilon^k\mu(B(x,r))~.$$ 
From \equ(lower) we obtain
$$\delta^{k2n}r^{2n}\le\varepsilon^k\mu(B(x,r))~,$$
implying the contradiction
$$2n\ge\frac{k\log \varepsilon +\log\mu(B(x,r))}{k\log\delta+\log r}
     @>>k\to\infty>\frac{\log\varepsilon}{\log\delta}\ge 3n~.$$
Hence $\por(\mu,x)\le\frac12$ for $\mu$-almost all $x\in\Bbb R^n$ giving 
the claim.

2. For sets it is obvious that if $\por(A)\ge p$ and $B\subset A$, then
$\por(B)\ge p$. The corresponding property holds for finite Radon measures:
if $B$ is a Borel set with $\mu(B)>0$, then 
$\por(\mu\vert_B,x)\ge\por(\mu,x)$ for $\mu$-almost all $x\in B$.
Indeed, according to the density point theorem \cite{M2, Corollary
2.14} we have for $\mu$-almost 
all $x\in B$ that $\mu(B(x,r)\cap B)\ge\frac12\mu(B(x,r))$ for all 
sufficiently small $r>0$. For all such $x$ and $r$ we have for all 
$\varepsilon,p>0$ and $z\in\Bbb R^n$ with $B(z,pr)\subset B(x,r)$ and 
$\mu(B(z,pr))\le\varepsilon\mu(B(x,r))$ that
$$\mu\vert_B(B(z,pr))\,\le\,\mu(B(z,pr))\le\varepsilon\mu(B(x,r))
   \le 2\varepsilon\mu\vert_B(B(x,r))~.$$
This implies the claim.

\medskip
 
We denote by $\spt(\mu)$ the support of $\mu$ which is the smallest closed set
such that the complement of it has $\mu$-measure zero.
Clearly  
$$
\por(\spt(\mu))\,\le\,\por(\mu)~.
$$ 
As illustrated by the following examples this inequality can be
strict. In fact, it is precisely this difference which makes the
definition of porosity important for physical measurements because it allows to
neglect systematically dust which is visible in porosities of sets
but not in those of measures.

\LIKEREMARK{Example 1}
Let $\delta_0$ be the Dirac measure at the origin, that is,
$\delta_0(A)=1$ if $0\in A$ and $\delta_0(A)=0$ if $0\notin A$. 
Let $\mu$ be the sum of $\delta_0$ and the Lebesgue measure $\Cal L^n$
restricted to $B(0,1)$, that is, $\mu=\delta_0+\Cal L^n|_{B(0,1)}$. Clearly
$\por(\mu,0)=\frac 12$ and $\por(\mu,x)=0$ for all $x\ne0$ with $|x|<1$.
Thus $\por(\mu)=\frac 12$. However, $\por(\spt(\mu))=\por(B(0,1))=0$.

\LIKEREMARK{Example 2}
Enumerate the rational numbers in the closed unit interval
$[0,1]$. Let $\delta _i$ be the Dirac measure on the $i^{\text{th}}$
rational point $x_i$.
Define $\mu=\sum_{i=1}^\infty 2^{-i} \delta _i$.
Then $\por(\mu,x_i)=\frac 12$ for all $i$ since for all $\epsilon $
there exists $r>0$ such that all the rationals in the $r$-neighbourhood
of $x_i$ have bigger index than $k+i$ for a fixed positive integer $k$
with $2^{-k}<\epsilon$. Hence $\por(\mu)=\frac 12$. 
Clearly $\por(\spt(\mu))=\por([0,1])=0$.

\medskip

For all $x\in\Bbb R^n$ and $r>0$ we have
$\lim_{\varepsilon\downarrow0}\por(\mu,x,r,\varepsilon)$
$=\por(\spt(\mu),x,r)$.
In particular,
$$\por(\spt(\mu),x)\,=\,\liminf_{r\downarrow0}\lim_{\epsilon \downarrow0} 
  \por(\mu,x,r,\varepsilon)~.$$
Thus, changing the order of taking the limits in \equ(porlocal) gives the
porosity of the support of the measure.

We will need later the following measurability property:

\REMARK
We will prove that for all $r>0$ and $\varepsilon>0$ the function
$x\mapsto\por(\mu,x,r,\varepsilon)$ is upper semi-continuous, that is,
$$\por(\mu,x,r,\varepsilon)\ge\limsup_{i\to\infty}
  \por(\mu,x_i,r,\varepsilon)\EQ(usc)$$
whenever $x_i\in\Bbb R^n$ are such that $\lim_{i\to\infty}x_i=x$.
We use the notations $p_i=\por(\mu,x_i,r,\varepsilon)$ and 
$p=\limsup_{i\to\infty}p_i$. Let $\delta>0$. For all $i$ there exists 
$z_i\in\Bbb R^n$ such that $B(z_i,(p_i-\frac\delta 2)r)\subset B(x_i,r)$
and 
$$\mu(B(z_i,(p_i-\frac\delta 2)r))\le\varepsilon\mu(B(x_i,r))~.\EQ(help)$$
By choosing $i$ so large that $\vert x-x_i\vert\le\frac{\delta r}2$ we have
$B(z_i,(p_i-\delta)r)\subset B(x,r)$. Further,
taking suitable subsequences we may assume that the sequence
$B(z_i,(p_i-\delta)r)$ converges with respect to the Hausdorff metric 
in the space of compact subsets of $\Bbb R^n$ (see \cite{R, Chapter 2.6})
and $p=\lim_{i\to\infty}p_i$. Then there is $z\in\Bbb R^n$ such that
$B(z,(p-2\delta)r)\subset\cap_iB(z_i,(p_i-\delta)r)\subset B(x,r)$.
Since the function $x\mapsto\mu(B(x,r))$ is upper semi-continuous
\cite{M2, Remark 2.10}, we obtain from \equ(help)
$$\varepsilon\mu(B(x,r))\ge\limsup_{i\to\infty}\varepsilon\mu(B(x_i,r))
  \ge\limsup_{i\to\infty}\mu(B(z_i,(p_i-\delta)r))
  \ge\mu(B(z,(p-2\delta)r))~.$$
Thus $\por(\mu,x,r,\varepsilon)\ge p-2\delta$. Since $\delta >0$ is
arbitrary,
this implies \equ(usc).

\medskip

We will consider the class of measures which satisfy the doubling condition:
\CLAIM Definition(doubling) A finite Borel measure $\mu$ on $\Bbb R^n$
satisfies the doubling condition at a point $x\in\Bbb R^n$ if
$$
\limsup_{r\downarrow0} {\mu(B(x,2r))\over \mu(B(x,r))}\,<\,\infty ~.
\EQ(20)
$$
We say that $\mu$ satisfies the doubling condition if \equ(20) holds for
$\mu$-almost all $x\in\Bbb R^n$.

\medskip

Expressing \clm(defpormu) in terms of porosities of sets, 
we will prove an analogue to \clm(mainset) for measures that
satisfy the doubling condition. We will also show that
the doubling condition is necessary for the validity of the relationship
between porosities of measures and sets.
Using \clm(mainset) we then obtain:

\CLAIM Theorem(main) 
There is a non-decreasing function $\Delta_n:[0,1/2]\to[0,1]$ satisfying
$$
\lim_{p\uparrow{1\over 2}} \Delta_n(p)\,=\,1
$$
such that
$$
\dimp(\mu) \,\le\, n-\Delta_n(\por(\mu))~\EQ(ineq)
$$
for all finite Borel measures $\mu$ on
$\Bbb R^n$ that satisfy the doubling condition.

\medskip

\REMARK 1. In \clm(main) one can take the same function $\Delta_n$ as 
in \clm(mainset).

2. From a practical point of view, the doubling condition is satisfied for 
recursively constructed physical measures. For example, in many physical 
applications there exist $a,b,s>0$ such that
$$ar^s\le\mu(B(x,r))\le br^s$$
for all $r>0$ and $x\in\spt(\mu)$ which clearly implies the validity
of the doubling condition.
\medskip

If the porosity of a measure $\mu$ which satisfies
the doubling condition is close to $\frac 12$, then according to \clm(main) 
the packing 
dimension of $\mu$ is not much bigger than $n-1$. One cannot expect that
small porosity implies big dimension. This is illustrated by the
following example.

\LIKEREMARK{Example 3} For all positive integers $k$ and $m$ there is a Borel 
probability measure $\mu$ on $\Bbb R$ such that $\dimp(\mu)=\frac 1k$ and
$\por(\mu)\le\frac 1m$. 

\LIKEREMARK{Construction}
Divide the closed unit interval $[0,1]$ into $m^k$ subintervals of length 
$m^{-k}$ and select $m$ of them by taking every $(m^{k-1})^{\text{th}}$ one.
Define a Borel probability measure $\mu_1$ by giving the same weight 
$\frac 1m$ to each of these intervals. Continue by dividing the selected
intervals into $m^k$ subintervals of length $m^{-2k}$ 
and choosing every $(m^{k-1})^{\text{th}}$ of them. Define a Borel probability
measure $\mu_2$ by attaching the weight 
$\frac 1{m^2}$ to each of these intervals and proceed in the same way. Then
$(\mu_i)$ converges weakly to a Borel probability measure $\mu$. Clearly
$\dimp(\mu)=\frac 1k$. It is not difficult to see that $\por(\mu)\le\frac 1m$. 
In fact, this construction is a simplified version of Example 4, and therefore 
we give no details here. 

\SECTION The proof of \clm(main)

Let $\mu$ be a finite Borel measure on $\Bbb R^n$.
In order to prove \clm(main) we first prove that if $\mu$ satisfies 
the doubling condition then 
$$
\beta(\mu)\,\ge\,\por(\mu)~,
\EQ(betaP)
$$
where
$$
\beta(\mu)\,=\,\sup\{\por(A): A\text { is a Borel set with }\mu(A)>0\}
$$
(see \cite{MM}).
We will obtain \clm(main) as a consequence of \equ(betaP) and \clm(mainset).
In Example 4 we will show that \equ(betaP) does not necessarily hold if the 
doubling condition is violated. That construction also indicates that the 
existence of the local dimension does not guarantee that \equ(betaP) holds.

Note that the inequality 
$$
\beta(\mu)\,\le\,\por(\mu)
\EQ(less)
$$
holds for any finite Radon measure $\mu$ on $\Bbb R^n$. In fact, if this is 
not the case, there exists $s$ such that $\por(\mu)<s<\beta (\mu)$. Using the 
density point theorem \cite{M2, Corollary 2.14}, we find a Borel set $A$ with 
$\mu(A)>0$ and $\por(A)>s$ such that
$$
\lim_{r\downarrow0}\frac{\mu(A\cap B(x,r))}
{\mu(B(x,r))}\,=\,1~
$$
for all $x\in A$. This means that for all $x\in A$ and $\epsilon>0$ we have
$\por(A,x,r)>s$ and
$$
\mu(A\cap B(x,r)) \,\ge\, (1-\epsilon )\mu(B(x,r))~\EQ(xx)
$$
for all sufficiently small $r>0$.
Hence for all such $r$
there exists $z\in\real^n$ with $B(z,sr)\subset B(x,r)\setminus A$.
By \equ(xx) this implies
$$
\mu(B(z,sr))\,\le\,\mu(B(x,r))-\mu(B(x,r)\cap A)\,\le\,\epsilon
\mu(B(x,r))~
$$
giving $\por(\mu)\ge s$. Thus \equ(less) holds.

While \equ(less) is valid without assuming the doubling condition,
it is needed for the opposite inequality:

\CLAIM Proposition(other) Let $\mu$ be a finite Borel measure on
$\real^n$. If $\mu$ satisfies the doubling condition,
then
$$
\beta(\mu)\,\ge\,\por(\mu)~.
$$
In particular, $\beta(\mu)=\por(\mu)$ for all finite Radon measures $\mu$
on $\Bbb R^n$ satisfying the doubling condition.

\PROOF Assume that $\beta(\mu)<\por(\mu)$. Let $s>0$ and $\delta>0$ be such
that $\beta(\mu)<s-\delta <s<\por(\mu)$. Setting  
$$
A\,=\, \{ x\in\spt(\mu) ~:~ \por(\mu,x) >s\}~,
$$
we have $\mu(A)>0$. 
Since $r\mapsto r\por(\mu,x,r,\varepsilon)$ is non-decreasing
and $r\mapsto\frac 1r$ is continuous, the lower limit in \equ(porlocal) does not 
change if $r$ is restricted to positive rationals. Also the limit as 
$\varepsilon$ goes to zero can be taken over rationals since 
$\liminf_{r\downarrow 0}\por(\mu,x,r,\varepsilon)$ is non-decreasing as
a function of $\varepsilon$. Thus by \equ(usc) the function 
$x\mapsto\por(\mu,x)$ is Borel measurable, and so $A$ is a Borel set.

For all positive and finite numbers $C$ define
$$
E_C\,=\, \{ x\in \spt(\mu)~:~ \mu(B(x,2r))>C\mu(B(x,r))\text{ for
some } r>0 \}~.
$$
Using the monotonicity of the mapping $r\mapsto\mu(B(x,r))$ it is easy to see
that the definition of $E_C$ is not altered if $r$ is restricted to positive
rationals. Therefore the Borel measurability of the mapping 
$x\mapsto\mu(B(x,r))$ \cite{M2, Remark 2.10} implies that $E_C$ 
is a Borel set for all $C$. 
Since $\mu$ satisfies the doubling condition,  
there is a positive and finite number $C$ such that
$\mu(E_C)\,<\,\frac{\mu(A)}2$. Hence
$\mu((\Bbb R^n\setminus E_C)\cap A)\,>\,\frac{\mu(A)}2 >0$.

Consider $x\in A$. For all sufficiently small $\epsilon >0$ and 
$r>0$ we have $\por(\mu,x,r,\epsilon )>s$. Hence for
all such $r$ and $\epsilon $, there is $z\in\real^n$ such that
$B(z,sr)\subset B(x,r)$ and $\mu(B(z,sr))\le\epsilon\mu(B(x,r))$.
We will prove that
$$
B(z,(s-\delta )r)\cap(\Bbb R^n\setminus E_C)\cap \spt(\mu) \,=\, \emptyset~.
\EQ(*)
$$
This gives the claim, since the fact that
$$ 
B(z,(s-\delta )r)\subset B(x,r)\setminus((\Bbb R^n\setminus E_C)\cap A
\cap \spt(\mu))~
$$
implies 
$$
\por((\Bbb R^n\setminus E_C)\cap A\cap\spt(\mu),x)\,\ge\, s-\delta ~
$$
giving $\beta(\mu)\ge s-\delta$ which is a contradiction.

To prove \equ(*), we assume that there exists 
$y\in B(z,(s-\delta )r)\cap(\Bbb R^n\setminus E_C)\cap \spt(\mu) $. 
Let $n$ be a positive integer such that $2^{-n+1}\le \delta$. Then
$$\align
   \mu(B(y,\delta r))&\,\le\,\mu(B(z,sr))\,\le\,\varepsilon\mu(B(x,r))
   \,\le\,\varepsilon\mu(B(y,2r))\cr
   &\,\le\,\varepsilon C^{n}\mu(B(y,2^{-n+1}r))
   \,\le\,\varepsilon C^{n}\mu(B(y,\delta r))~.
\endalign$$
This gives a contradiction because we can choose 
$\varepsilon$ as small as we wish.
\qed

\medskip

Using \equ(PD3) and \clm(other) we can estimate both the packing dimensions
and porosities of measures satisfying the doubling condition
in terms of corresponding quantities of sets.
This gives an easy way to prove \clm(main) using \clm(mainset):

\medskip

\noindent\bf{Proof of \clm(main).\/}\rm\quad
Let $\Delta_n:[0,\frac 12]\to[0,1]$ be as in \clm(mainset).
Consider a finite Borel measure $\mu$ on $\Bbb R^n$ which satisfies
the doubling condition. Since $\beta(\mu)\ge\por(\mu)$ by \clm(other), 
we find for all $\delta>0$ a Borel set
$A\subset\Bbb R^n$ with $\mu(A)>0$ such that 
$\por(A)\ge\por(\mu)-\delta$.
Now \equ(PD3) and \clm(mainset) give
$$
\dimp(\mu)\le\dimp(A)\le n-\Delta_n(\por(A))\le n-\Delta_n(\por(\mu)-\delta).
$$
The claim follows using the continuity of the function $\Delta_n$.
\qed 

\SECTION The role of the doubling condition

In this section we show that \clm(other) is not generally valid unless
the measure $\mu$ satisfies the doubling condition.  

\LIKEREMARK {Example 4} There exists a Borel probability measure $\mu$ on
$\Bbb R$ with the following properties:
$$\align
 &\beta(\mu)\,=\,0~,\NR(1)
 &\por(\mu)\,=\,\frac 13~,\NR(2)
 &\mu\Big(\Big\{x:\limsup_{r\downarrow 0}
  \frac{\mu(B(x,2r))}{\mu(B(x,r))}<\infty\Big\}\Big)\,=\,0,\text{ and }~\NR(3)
 &\dimp(\mu)\,=\,0~.
\NR(4) 
\endalign
$$

\LIKEREMARK{Construction} 
For all $i=1,2,\dots$ we first define a Borel probability measure $\mu_i$
such that its restriction to any closed dyadic subinterval of the closed 
unit interval of length $2^{-i}$ is a constant multiple of Lebesgue measure.
For $i=1,2,\dots$ let $J_i$ be the set of all
$i$-term sequences of integers 0 and 1 and let $J_\infty$ be the corresponding
set of infinite sequences, that is,
$$J_i=\{(j_1,j_2,\dots,j_i):j_m\in\{0,1\}\text{ for all }m=1,\dots,i\}$$
and
$$J_\infty=\{(j_1,j_2,\dots):j_m\in\{0,1\}\text{ for all }m=1,2,\dots\}.$$
We denote by $I_{j_1\dots j_i}$ the closed dyadic interval
of length $2^{-i}$ whose left endpoint in binary representation
is $0,j_1j_2\cdots j_i$.
Let $(p_i)$, $0<p_i<1$, be a decreasing sequence of real numbers tending to 
zero. The measure $\mu_i$ is defined by requiring that 
$$
\mu_i(I_{j_1\dots j_i})\,=\,\prod_{k=1}^i (1-p_k)^{j_k} p_k^{1-j_k}
$$
for all $(j_1,\dots,j_i)\in J_i$.
It is easy to see that $(\mu_i)$ converges weakly to
a Borel probability measure $\mu$ such that $\spt(\mu)=[0,1]$. 

Equivalently one can think of the measure $\mu$ as the projection of 
a natural product measure on the code space. In fact, defining
$\nu_k(\{0\})=p_k$ and $\nu_k(\{1\})=1-p_k$ for all $k=1,2,\dots$, the product
measure $\prod_{k=1}^\infty\nu_k$ is a Borel probability measure on the
code space $J_\infty$ (equipped with the product topology)  and the measure 
$\mu$ is its image under the projection $\pi:J_\infty\to[0,1]$, where 
$\pi((j_1,j_2,\dots))=\sum_{m=1}^\infty j_m2^{-m}$, that is, the binary
representation of a point in $[0,1]$.

\medskip

The measure $\mu$ has the following property:

\medskip
 
\CLAIM Lemma(LemmaC) Let $0<\delta<1$. Given any $\varepsilon>0$ the
following property holds for all sufficiently large positive integers  $k$:
for all closed dyadic subintervals $[x_1,x_2]$ of the unit interval $[0,1]$
of length $2^{-k}$ we have
$$
\mu([x_1,x_2-2^{-k}\delta])\,\le\,\varepsilon\mu([x_1,x_2])~.
\EQ(muprop)
$$

\noindent\bf{Proof of \clm(LemmaC).}
\rm
Consider the positive integer $\ell$ such that 
$2^{-\ell}\le\delta<2^{-\ell+1}$.
Since the interval $[x_1,x_2-2^{-k}\delta]$ is contained in the union of
$2^{\ell}-1$ closed dyadic subintervals of $[x_1,x_2]$ of length 
$2^{-k-\ell}$ and of measure at most $p_{k+1}\mu([x_1,x_2])$ (we take
all subintervals of $[x_1,x_2]$ of length $2^{-k-\ell}$ 
except the right most one as covering sets) we have
$$\mu([x_1,x_2-2^{-k}\delta])\le(2^{\ell}-1)p_{k+1}\mu([x_1,x_2]).$$
Choosing $k$ so large that $(2^{\ell}-1)p_{k+1}\le\varepsilon$
gives the claim. 
\qed

\medskip

\clm(LemmaC) is essential when proving properties \equ(1) -- \equ(4):

\medskip 

\LIKEREMARK{Proof of properties \equ(1) -- \equ(4)}
For \equ(1) we assume that $\beta(\mu)>0$. Then there exist
a positive integer $k$, a real number $R$ with $0<R<1$, and
$E\subset[0,1]$ with $\mu(E)>0$ such that for all $x\in E$ we have
$$\por(E,x,r)\ge 2^{-k}\EQ(+)$$
for all $0<r\le R$. Set $N=2^{k+4}$. Let $i_0$ be a
positive integer with $2^{-i_0}\le2^{-k-2}R$.
We will first show that if $i$ is a positive integer with
$i\ge i_0$, then, given any family
$\{D_1,\dots,D_N\}$ of successive closed
dyadic subintervals of $[0,1]$ of length
$2^{-i}$, there is $1\le j\le N$ such that
$$E\cap D_j=\emptyset.\EQ(**)$$
If this were not the case, then $D_j\cap E\ne\emptyset$ for all $j=1,\dots,N$.
Let $M=N/2$. Consider $x\in E\cap D_M$
and set $r_i=2^{2+k-i}$. Denote by $d_1$ the left-hand end-point of $D_1$
and  by $d_N$ the right-hand end-point of $D_N$. Now 
$|x-d_1|\ge(M-1)2^{-i}$, $|x-d_N|\ge(M-1)2^{-i}$, and
$(M-1)2^{-i}\ge N2^{-i-2}=r_i$, and therefore we obtain
$$B(x,r_i)\subset\bigcup_{j=1}^ND_j.$$
Further, since $2\cdot2^{-i}<r_i\le R$ and all dyadic intervals $D_j$ meet $E$,
we have
$$\por(E,x,r_i)\le\frac{2^{-i}}{r_i}=2^{-k-2}$$
which contradicts \equ(+). Thus \equ(**) holds.

We complete the proof of \equ(1) by showing that the property \equ(**) 
implies that $\mu(E)=0$. Set $\ell=k+4$. We may assume that $i_0=m\ell$ for 
some $m\in\Bbb N$. Denote by $F$ the set of numbers in $[0,1]$ 
whose base two expansion does not contain the sequence 
$j_{n\ell}=0,j_{n\ell+1}=0,\dots,j_{(n+1)\ell-1}=0$ for any integer $n\ge m$. 
Let $i\ge i_0$. An $N$-block at stage $i$ is a family of $N$ successive closed
dyadic subintervals of $[0,1]$ of length $2^{-i}$ which belong to the same 
dyadic interval of length $2^{-i+\ell}$ at stage $i-\ell$. By \equ(**) 
in each of these 
$N$-blocks there is at least one interval which 
does not intersect $E$. Since the left-most interval of each $N$-block 
has the smallest measure, we have $\mu(E)\le\mu(F)$. Further,
choosing for all $i$
$$p_i=\frac 1{\log(i+2)}$$ 
we have
$$\eqalign
 {\mu(F)&\le\prod_{j=0}^\infty\Big(1-p_{j\ell+i_0}
  \cdot\dots\cdot p_{(j+1)\ell+i_0-1}\Big)\cr
  &=\exp\Big(\sum_{j=0}^\infty\log(1-p_{j\ell+i_0}\cdot\dots\cdot 
   p_{(j+1)\ell+i_0-1})\Big)\cr
  &\le\exp\Big(-\frac12\sum_{j=0}^\infty p_{j\ell+i_0}\cdot\dots\cdot 
   p_{(j+1)\ell+i_0-1}\Big)\cr
  &\le\exp\Big(-\frac12\sum_{j=0}^\infty
   \frac1{(\log((j+1)\ell+i_0+1))^{\ell}}\Big)=0.\cr}\EQ(kuku)
$$
Hence \equ(1) holds.

In order to prove \equ(2) let $x\in [0,1]$ and $r>0$. 
Consider a positive integer $i$ such that $2^{-i}\le r< 2^{-i+1}$.
Let $D_i$ be a closed dyadic subinterval of $[0,1]$ of length
$2^{-i+1}$ which contains $x$. 
We denote by $D_i^\L$ and $D_i^\R$ the neighbouring closed
dyadic intervals of $D_i$
of length $2^{-i+1}$ situated on left and right, respectively.
The interval $D_i$ is the union of
four closed dyadic intervals of length $2^{-i-1}$. 
Let $a_i$, $b_i$, $c_i$, $d_i$, and $e_i$ be the end-points of 
these four intervals from left to right (see \fig(fig.pps)).
Then 
$$\eqalign
{\mu([a_i,b_i])\,&=\,p_ip_{i+1}\,\mu(D_i)~,\cr
\mu([b_i,c_i])\,&=\,p_i(1-p_{i+1})\,\mu(D_i)~,\cr
\mu([c_i,d_i])\,&=\,(1-p_i)p_{i+1}\,\mu(D_i)~,\cr
\mu([d_i,e_i])\,&=\,(1-p_i)(1-p_{i+1})\,\mu(D_i)~,\cr
}
\EQ(**a)
$$ 
giving
$$\eqalign{
\frac{p_i^2}2\mu(D_i)\,&\,\le\,\,\mu([a_i,b_i])\,\le\, p_i^2\mu(D_i)~,\cr
\frac {p_i}2\mu(D_i)\,&\le\,\mu([b_i,c_i])\,\le\, p_i\mu(D_i)~,\cr
\frac {p_i}4\mu(D_i)\,&\le\,\mu([c_i,d_i])\,\le\, p_{i}\mu(D_i)~,\cr
\frac 14\mu(D_i)\,&\le\,\mu([d_i,e_i])\,\le\,\mu(D_i)~.\cr
}
\EQ(***)
$$
(We can assume that $i$ is big enough such that $p_i<\frac 12$.)
The following concept of scaling is used to describe the behaviour 
of measures of dyadic intervals.
If $D\subset[0,1]$ is a dyadic interval of length $2^{-i-1}$ we say
that $\mu(D)$ scales like $p_i^k$ for some integer $k$ if
there is a constant $c$ independent of $i$ such that
$$\frac 1cp_i^k\mu(D_i)\le\mu(D)\le cp_i^k\mu(D_i).$$ 
In particular,
\equ(***) implies that $\mu([a_i,b_i])$ scales like $p_i^2$, 
$\mu([b_i,c_i])$ scales 
like $p_i$, $\mu([c_i,d_i])$ scales like $p_i$, 
and $\mu([d_i,e_i])$ scales like 1 (see \fig(fig.pps)).

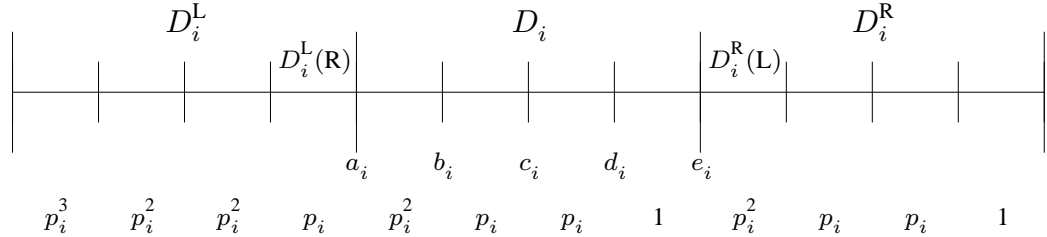
\figurewithtexplus fig.pps fig.tex 18 15 -14 The scaling properties of 
the intervals.\cr

We denote by $D_i^\R(\L)$ the left-most closed subinterval of
$D_i^\R$ of length $2^{-i-1}$, and by $D_i^\L(\R)$ the right-most closed
subinterval of $D_i^\L$ of length $2^{-i-1}$ (see \fig(fig.pps)).
The length of the shortest possible dyadic interval containing either
both $[a_i,b_i]$ and $D_i^\L(\R)$ or both $[d_i,e_i]$ and $D_i^\R(\L)$ 
is at least $2^{-i+2}$. Let $D$ be the shortest dyadic interval containing
$[d_i,e_i]$ and $D_i^\R(\L)$ and let $2^{-m}$, $m\le i-2$, be its length.
Then $[d_i,e_i]$ is reached from $D$ after stepping left at stage
$m+1$ and then always right, and $D_i^\R(\L)$ is reached after stepping 
first right at stage $m+1$ and after that always left, and so
$$\frac{\mu([d_i,e_i])}{\mu(D_i^\R(\L))}
  =\frac {p_{m+1}(1-p_{m+2})\cdot\dots\cdot(1-p_{i+1})}
        {(1-p_{m+1})p_{m+2}\cdot\dots\cdot p_{i+1}}.\EQ(?)$$
Similarly
$$\frac{\mu(D_i^\L(\R))}{\mu([a_i,b_i])}
  =\frac {p_{n+1}(1-p_{n+2})\cdot\dots\cdot(1-p_{i+1})}
        {(1-p_{n+1})p_{n+2}\cdot\dots\cdot p_{i+1}},\EQ(!)$$
where $n\le i-2$ is the biggest possible stage where $D_i^\L(\R)$ and 
$[a_i,b_i]$ belong to the same dyadic interval.

We consider the case where we have the minimum relative weight for $D_i^\L(\R)$.
Other cases can be treated similarly. 
By \equ(!) the relative weight of $D_i^\L(\R)$ obtains the minimum for $n=i-2$. 
This means that
$D_i^\L(\R)$ and $[a_i,b_i]$ belong to the same dyadic interval of 
length $2^{-i+2}$. Then \equ(!) implies 
$$\mu(D_i^\L(\R))=\frac {p_{i-1}(1-p_i)(1-p_{i+1})}
        {(1-p_{i-1})p_ip_{i+1}}\mu([a_i,b_i])$$
giving 
$$\frac 14p_i\mu(D_i)\le\mu(D_i^\L(\R))\le 4p_i\mu(D_i).$$
Hence $\mu(D_i^\L(\R))$ scales like $p_i$. 
As before we see that the other three closed
dyadic subintervals of $D_i^\L$ of length
$2^{-i-1}$ scale like $p_i^3$, $p_i^2$, and $p_i^2$ from left to right
(see \fig(fig.pps)).

Since $D_i^\L(\R)$ and $[a_i,b_i]$ belong to the same 
dyadic interval at stage $i-2$, the stage $m$ where $[d_i,e_i]$ 
and $D_i^\R(\L)$
belong to the same dyadic interval cannot be bigger than $i-3$.   
Here we consider the case $m=i-3$. Again other cases are similar to this one. 
Using \equ(?) for $m=i-3$ we obtain the maximum value for the relative
weight of $D_i^\R(\L)$
$$\mu(D_i^\R(\L))=\frac{(1-p_{i-2})p_{i-1}p_ip_{i+1}} 
             {p_{i-2}(1-p_{i-1})(1-p_i)(1-p_{i+1})}\mu([d_i,e_i]).$$
This implies that the closed dyadic subintervals of $D_i^\R$ of length
$2^{-i-1}$ scale like $p_i^2$, $p_i$, $p_i$, and 1 (see \fig(fig.pps)).

Let $\delta$, $\varepsilon>0$. We may assume that $i$ is so
large, that is, $[a_i,e_i]$ is so short, that \clm(LemmaC) holds.
Assume first that $a_i\in B(x,r)$. Since $r\ge 2^{-i}$ 
and $x\in [a_i,e_i]$, we obtain 
$[a_i,c_i]\subset B(x,r)$. By \clm(LemmaC) we have  
$\mu([a_i,c_i-2^{-i}\delta])$
$\le\varepsilon\mu([a_i,c_i])\le\varepsilon\mu(B(x,r))$. 
If $B(x,r)\subset(a_i-2^{-i},e_i)$, then $r\le 3\cdot2^{-i-1}$, and so
$$\por(\mu,x,r,\varepsilon)\ge
  \frac{\frac12(c_i-a_i-2^{-i}\delta)}{3\cdot2^{-i-1}}
  =\frac 13(1-\delta).\EQ(E1)$$
If $B(x,r)$ is not contained in $(a_i-2^{-i},e_i)$, then either $e_i\in B(x,r)$
or $a_i-2^{-i}\in B(x,r)$. Consider first the case where $e_i\in B(x,r)$. Then
$[a_i,e_i]\subset B(x,r)$. Thus \clm(LemmaC) implies that
$\mu([a_i,e_i-2^{-i+1}\delta])\le\varepsilon\mu(B(x,r))$ giving
$$\por(\mu,x,r,\varepsilon)\ge
  \frac{\frac12(e_i-a_i-2^{-i+1}\delta)}{2^{-i+1}}
=\frac 12(1-\delta).\EQ(E2)$$ 
In the case where $a_i-2^{-i}\in B(x,r)$ 
and $e_i\notin B(x,r)$ we have 
$$\frac{x-a_i}r\le\min\Big\{1-\frac{2^{-i}}r, 
\frac{2^{-i+1}}r-1\Big\}\le\frac13.\EQ(M)$$
Using \clm(LemmaC) and the previously mentioned 
scaling properties of $\mu$ we find 
a constant $C$ independent of $i$ and $\varepsilon$ such that
$\mu([x-r,a_i-2^{-i+1}\delta])\le C\varepsilon\mu(B(x,r))$. From \equ(M) it 
follows that
$$\por(\mu,x,r,C\varepsilon)
  \ge\frac{\frac12(a_i-2^{-i+1}\delta-x+r)}r 
  \ge\frac12-\frac{x-a_i}{2r}-\frac{2^{-i}\delta}{r}\ge
  \frac13-\delta.\EQ(E3)$$
Finally we consider the remaining case where $a_i\notin B(x,r)$. 
Then $e_i\in B(x,r)$, and so by \clm(LemmaC) and the scaling properties of
$\mu$ there is a constant $C$ independent of $i$ and $\varepsilon$
such that $\mu([x-r,e_i-2^{-i+1}\delta])\le C\varepsilon\mu(B(x,r))$. 
Since $x\le e_i$ we obtain  
$$\por(\mu,x,r,C\varepsilon)\ge\frac{\frac12(e_i-2^{-i+1}\delta-x+r)}r
  \ge\frac{\frac12(r-2^{-i+1}\delta)}r\ge\frac12-\delta.\EQ(E4)$$
By \equ(E1) -- \equ(E4) we have $\por(\mu,x)\ge\frac13$ for all $x\in[0,1]$, 
and so $\por(\mu)\ge\frac13$. 

For the opposite inequality consider a sequence $(\delta_k)$ of positive real
numbers tending to zero. Then for all $k$ there is a positive integer
$m_k$ and a sequence $(1,0,0,1,1,\dots,1,1)\in J_{m_k}$ such that
if the base two expansion of a point $x\in D_i$ contains this sequence   
from the $(i-1)^{\text{th}}$ place, then $D_i^\L$ and $D_i$ belong to the same
dyadic subinterval of $[0,1]$ of length $2^{-i-2}$ and
$0\le b_i-x\le2^{-i-1}\delta_k$.
For all $k$ let $A_k$ be the set of points $x\in[0,1]$ whose base two 
expansion contains the sequence $(1,0,0,1,1,\dots,1,1)\in J_{m_k}$
infinitely many times and let
$$B_k=\{x\in[0,1]: \por(\mu,x)\le\frac13+\delta_k\}.$$
Then $A_k\subset B_k$ for all $k$. In fact, as in \equ(E1) we see that for all
$\varepsilon>0$ and for all $x\in A_k$
$$\por(\mu,x,2^{-i-1}(3-\delta_k),\varepsilon)
  \le\frac{\frac12(c_i-a_i)}{2^{-i-1}(3-\delta_k)}\le\frac13+\delta_k$$
for all $i$ large enough implying that $x\in B_k$.
Further, as in \equ(kuku) we obtain that $\mu(A_k)=1$ for all $k$, giving
$\mu(\cap_{k=1}^\infty B_k)=1$. Hence \equ(2) holds. 

As in \equ(kuku) it can be shown
that $\mu$-almost every point has the sequence $01$
infinitely many times in its expansion. Thus $\mu$-almost every point 
belongs for arbitrarily large positive integers $i$ to the second one 
of the four dyadic subintervals of a dyadic interval of length $2^{-i+1}$.
This implies that
$\mu(B(x,2^{-i-1}))$ scales like $p_i$ and $\mu(B(x,3\cdot 2^{-i-1}))$ 
scales like 1 giving
$$\limsup_{r\downarrow 0}\frac{\mu(B(x,3r))}{\mu(B(x,r))}=\infty$$
for $\mu$-almost all $x\in [0,1]$. Thus \equ(3) is proved. 

It remains to show that \equ(4) holds. 
By \cite{C, Lemma 2.3} it is enough to prove that 
for $\mu$-almost all $x\in [0,1]$
$$\lim_{i\to\infty}\frac 1i\log\mu(I_{j_1\dots j_i}(x))=0,$$
where $I_{j_1\dots j_i}(x)$ is the dyadic
subinterval of $[0,1]$ of length $2^{-i}$ which contains $x$. Note that
$$\frac 1i\log\mu(I_{j_1\dots j_i}(x))
   =\frac 1i\sum_{m=1}^i\big(\delta_{j_m,0}\log p_m+\delta_{j_m,1}
   \log(1-p_m)\big)=:A_i,$$
where $\delta_{j,k}=1$ if $j=k$ and $\delta_{j,k}=0$ if $j\ne k$.
Let $Y_i$ be a random variable such that $Y_i=\log p_i$ with probability $p_i$
and $Y_i=\log(1-p_i)$ with probability $1-p_i$. 
Then the expectation of $Y_i$ is
$$E_i=p_i\log p_i+(1-p_i)\log(1-p_i).$$
Clearly the variance 
$$V_i=p_i(\log p_i)^2+(1-p_i)(\log(1-p_i))^2
      -\big(p_i\log p_i+(1-p_i)\log(1-p_i)\big)^2$$
goes to zero as $i$ tends to infinity, and
so there exists a constant $C$ such that
$|V_i|\le C$ for all $i$. According to Kolmogorov's Criterion 
\cite{Fe, (X.7.2)} the strong law of large numbers is valid 
\cite{Fe, (X.7.1)}, that is, for $\mu$-almost all $x\in [0,1]$
$$\lim_{i\to\infty}A_i=\lim_{i\to\infty}\frac1i\sum_{m=1}^i E_m.$$
Since $|(1-p_m)\log(1-p_m)|\le p_m$ and the sums 
$\frac 1i\sum_{m=1}^i p_m$ and $\frac 1i\sum_{m=1}^i p_m\log p_m$
go to zero as $i$ goes to infinity, we obtain the claim.
\qed

\REMARK
For all $A\subset\Bbb R^n$ define
$$\overline\por(A)=\inf\{\overline\por(A,x):x\in A\}$$
where
$$\overline\por(A,x)=\limsup_{r\downarrow0}\por(A,x,r).$$
For all finite Borel measures $\mu$ on $\Bbb R^n$ set
$$\overline\beta(\mu)=\sup\{\overline\por(A): A\text{ is a Borel set with }
  \mu(A)>0\}.$$
According to \cite{MM, Theorem 1.1} the measure $\mu$ satisfies
the doubling condition if $\overline\beta(\mu)<\frac 12$. 
Example 4 shows that the assumption $\underline\beta(\mu)<\frac 12$ 
does not necessarily guarantee this.

\SECTION One dimensional case

In this section we study the situation in $\Bbb R$. 
By considering the class of strongly porous measures
we prove that the 
doubling condition (\clm(doubling)) is not necessary for the validity of 
\clm(main) although without it \clm(other) is not true. 
 
\CLAIM Definition(uni) Let $\mu$ be a finite Borel measure on $\Bbb R^n$. We
say that $\mu$ is uniformly $p$-porous if for all $\varepsilon>0$ there
exists $R_\varepsilon>0$ such that for $\mu$-almost all $x\in\spt(\mu)$
$$\por(\mu,x,r,\varepsilon)\ge p\EQ(uporo)$$
for all $0<r\le R_\varepsilon$. Further, $\mu$ is called strongly $p$-porous
if $\por(\mu)\ge p$ and if the following property holds 
for all $q<p$: given any Borel set $A\subset\Bbb R^n$ with
$\mu(A)>0$ such that $\por(\mu,x)>q$ for all $x\in A$ 
there exists a Borel set
$B\subset A$ with $\mu(B)>0$ such that $\mu\vert_B$ is uniformly $q$-porous.

\REMARK 1. The upper semi-continuity of the function 
$x\mapsto\por(\mu,x,r,\varepsilon)$ implies that if \equ(uporo) is true for 
$\mu$-almost all $x\in\spt(\mu)$ then it is true for all $x\in\spt(\mu)$.

2. We showed in Remark 2 after \clm(defpormu) that
the restriction of 
a Radon measure to a $p$-porous Borel set is $p$-porous. However, that 
argument does not imply that the restriction to a uniformly $p$-porous Borel 
set would yield a uniformly $p$-porous measure. 

\medskip

\CLAIM Proposition(onemain) 
There is a non-decreasing function $d:[0,1/2]\to[0,1]$ satisfying
$$\lim_{p\uparrow\frac12}d(p)=1$$
such that
$$\dimH(\mu)\le1-d(p)$$
for all finite strongly $p$-porous Borel measures $\mu$ on $\Bbb R$~.

\PROOF We may assume that $\dimH(\mu)>0$. Since $\por(\mu)\ge p$, given any 
$q<p$, there exists a Borel set $A$ with $\mu(A)>0$ such that 
$\por(\mu,x)>q$ for all
$x\in A$. Let $0<s<\dimH(\mu)$. Then by \equ(HD) there are
$R>0$ and a Borel set $E\subset A$ with $\mu(E)>0$ such that 
$$\mu(B(x,r))\le r^s \EQ(ups)$$ 
for all $x\in E$ and for all $0<r<R$. 
Using \clm(uni) we find a Borel set $B\subset E$ with $\mu(B)>0$ such that 
$\nu=\mu\vert_B$ is uniformly $q$-porous. In particular, 
$$\nu(B(x,r))\le r^s\EQ(upnu)$$
for all $x\in\spt(\nu)$ and $0<r<R$.

Intuitively our argument below is based on the fact that if the porosity is 
close to $\frac 12$ then for all sufficiently small $r>0$ 
there exists an interval of length close to $r$ inside $B(x,r)$
for $x\in\spt(\nu)$ such
that the measure of this interval is close to $\nu(B(x,r))$. Iterating 
this we find a ball which has a very small radius compared to $r$ and which
has measure quite close to $\nu(B(x,r))$. 
This forces the dimension to be small. 

We may assume that $\nu$ is non-atomic since otherwise 
$\dimH(\mu)\le\dimH(\nu)=0$. Assume that $\frac{63}{128}<q<p$.
Let $0<\delta<\frac 1{64}$ with $q'=\frac{1-\delta}2<q<p$. 
Consider $0<\varepsilon<\frac{\delta}2$. Let $R_\varepsilon$ be such that
$$\por(\nu,x,r,\varepsilon)\ge q \EQ(nuporo)$$
for all $0<r\le R_\varepsilon$ and for all $x\in\spt(\nu)$.
Let $n$ be the biggest integer such that $2^{-n(n+1)}>\delta$. 

\medskip

The following lemma is essential in our proof:

\CLAIM Lemma(essent)
Let $a<b<c$ be real numbers such that $c-b\le R_\varepsilon$, 
$b-a\ge\frac{1-\delta}{1+\delta}(c-b)$ 
and $\nu([a,b])\le\nu([b,c])$.
Then one of the following properties holds:
$$\align
&(1)\text{ There is }z'\in\spt(\nu)\cap[b,c]\text{ with }
 \nu(B(z',2\delta(c-b)))\ge2^{-n}(1-5\varepsilon)\nu([b,c]).\\
&(2)\text{ There are }b\le a'<b'<c'\le c\text{ such that }
 b'-a'\ge\frac {1-\delta}{1+\delta}(c'-b'),\\
&\phantom{(2)a}c'-b'\le\frac12(1+\delta)(c-b),\quad
 \nu([a',b'])\le\nu([b',c']),\\
&\phantom{(2)a}\text{and }
 \nu([b',c'])\ge(1-2^{-n})(1-5\varepsilon)\nu([b,c]). 
\endalign$$

\REMARK Note that the choice of $n$ guarantees that in the case
(1) we gain much more in the radius than we loose in the weight.

\noindent{\bf Proof of \clm(essent).}
Since $\nu$ has no atoms there exists $y\in\spt(\nu)\cap(b,c)$ such that
$$2\varepsilon\nu([b,c])<\nu([b,y])<3\varepsilon\nu([b,c]). \EQ(start)$$
This gives $\nu([y,c])\ge(1-3\varepsilon)\nu([b,c])$. 
Therefore the requirement $c-y\le2\delta(c-b)$ implies (1) with $z'=y$. 
Thus we may now assume that $c-y>2\delta(c-b)$. If
$\nu([y+2\delta(c-b),c])=0$, then (1) holds again  because by \equ(start) we 
obtain $\nu(B(y,2\delta(c-b)))\ge\nu([y,y+2\delta(c-b)])
\ge (1-3\varepsilon)\nu([b,c])$.
 
It remains to consider the case $c-y>2\delta(c-b)$ and
$\nu([y+2\delta(c-b),c])\ne0$. Let 
$y'=\inf\{\spt(\nu)\cap[y+2\delta(c-b),c)\}$. Suppose first that 
$y'\ge b+\frac12(1-\delta)(c-b)$ (see \fig(f1.pps)).
\figurewithtexplus f1.pps f1.tex 18 16 -15 The case
$y'\ge b+\frac 12(1-\delta)(c-b)$.\cr
\noindent Setting $I_1=[y,y+2\delta(c-b)]$ and $I_2=[y',c]$ we conclude from 
\equ(start) that
$$\nu(I_1)+\nu(I_2)=\nu([b,c])-\nu([b,y])\ge(1-3\varepsilon)\nu([b,c]).$$
Note that (1) holds in the case when
$\nu(I_1)\ge 2^{-n}(1-3\varepsilon)\nu([b,c])$. If the opposite
inequality is valid, we have 
$\nu(I_2)>(1-2^{-n})(1-3\varepsilon)\nu([b,c])$ giving (2). (To check this
choose $a'=b$, $b'=y'$, and $c'=c$.)

We are left with the case $y'<b+\frac 12(1-\delta)(c-b)$.
Using the fact that $\nu$ is uniformly $q$-porous, we find
$z\in\Bbb R$ such that $B(z,q'(c-y'))\subset B(y',c-y')$ and 
$\nu(B(z,q'(c-y')))\le\varepsilon\nu(B(y',c-y'))$
giving
$$\nu(B(z,q'(c-y')))\le2\varepsilon\nu([b,c])~,\EQ(small)$$ 
since $B(y',c-y')\subset [a,c]$ and $\nu([a,b])\le \nu([b,c])$.
From \equ(start) and \equ(small) we get $[b,y]\not\subset B(z,q'(c-y'))$ and
claim that
$$B(z,q'(c-y'))\subset [b,c]~. \EQ(sub)$$
(The possibility which is excluded here is that the whole ball is to the 
left of $y$ (see \fig(f2.pps)).)
This being not the case gives $z-q'(c-y')<b$. Since
$y-(y'-(c-y'))\le c-y'-2\delta(c-b)< 2q'(c-y')$
we have $B(z,q'(c-y'))\not\subset [y'-(c-y'),y]$, giving $y<z+q'(c-y')$. 
This implies that $[b,y]\subset B(z,q'(c-y'))$ which is a 
contradiction.

Now we split our study into three cases depending on the positions of
$[y,y']$ and $B(z,q'(c-y'))$. First assume that $[y,y']\subset
B(z,q'(c-y'))$ as in \fig(f2.pps).
\figurewithtexplus f2.pps f2.tex 16 16 -12 The case
$[y,y']\subset B(z,q'(c-y'))$.\cr 
\noindent Since $y'<b+\frac12(1-\delta)(c-b)$, we have 
$b+2q'(c-y')\ge b+\frac12(1-\delta)(c-b)$,
and so we obtain from \equ(sub) that 
$[y,b+\frac12(1-\delta)(c-b)]\subset B(z,q'(c-y'))$.
Hence by \equ(start) and \equ(small)
$$\nu([b+\frac12(1-\delta)(c-b),c])\ge\nu([b,c])-\nu([b,y])-\nu(B(z,q'(c-y')))
  \ge(1-5\varepsilon)\nu([b,c])$$
implying (2). (To verify this choose $a'=b$, $b'=b+\frac12(1-\delta)(c-b)$, 
and $c'=c$.)

Next assume that $[y,y']\not\subset B(z,q'(c-y'))$ and $y'\in
B(z,q'(c-y'))$ as in \fig(f3.pps).
\figurewithtexplus f3.pps f3.tex 18 16 -15 The case
$[y,y']\not\subset B(z,q'(c-y'))$ and $y'\in B(z,q'(c-y'))$.\cr 
\noindent Then the measure 
$$\nu([b,c])-\nu([b,y])-\nu(B(z,q'(c-y')))\ge (1-5\varepsilon)\nu([b,c])$$ 
is divided between two disjoint intervals 
$I_3=[y,\min\{z-q'(c-y'), y+2\delta(c-b)\}]$ and 
$I_4=[z+q'(c-y'),c]$ contained in  
$[y,y+2\delta(c-b)]$ and $[b+\frac12(1-\delta)(c-b),c]$, respectively. 
If $\nu(I_3)\ge2^{-n}(1-5\varepsilon)\nu([b,c])$, then (1) holds. 
When $\nu(I_4)\ge(1-2^{-n})(1-5\varepsilon)\nu([b,c])$ we obtain (2).
(To check this choose
$a'=z-q'(c-y')$, $b'=z+q'(c-y')$, and $c'=c$.)

In the remaining case we have
$[y,y']\not\subset B(z,q'(c-y'))$ and $y'\notin B(z,q'(c-y'))$
giving $y'<z-q'(c-y')$ since 
$b+2q'(c-y')\ge b+\frac12(1-\delta)(c-b)>y'$ as in \fig(f4.pps).
\figurewithtexplus f4.pps f4.tex 17 16 -14 The case
$[y,y']\not\subset B(z,q'(c-y'))$ and $y'\notin B(z,q'(c-y'))$.\cr 
\noindent Note that
$c-y'-2q'(c-y')=\delta(c-y')\le\delta(c-b)$ which means that the set 
$[y',c]\setminus B(z,q'(c-y'))$ is the union of at most two intervals of
length at most $\delta(c-b)$, and so the measure 
$$\nu([b,c])-\nu([b,y])-\nu(B(z,q'(c-y')))\ge (1-5\varepsilon)\nu([b,c])$$ 
is divided between at most three intervals of length at most $2\delta(c-b)$.
(Take the two above intervals and $[y,y+2\delta(c-b)]$ which has the same 
$\nu$-measure as $[y,y']$.) Hence there exists 
an interval of length at most $2\delta(c-b)$
having $\nu$-measure at least $\frac 13(1-5\varepsilon)\nu([b,c])$, and so
(1) is satisfied.
\qed

\medskip

\noindent{\bf The continuation of the proof of \clm(onemain).}
Let $x\in\spt(\nu)$ and $0<r<\min\{R,\frac{R_\varepsilon}2\}$.
Since $\nu$ is uniformly $q$-porous, $B(x,r)$ contains 
an interval $[a,a+2q'r]$ such that
$\nu([a,a+2q'r])\le\varepsilon\nu(B(x,r))$. 
Hence either $\nu([x-r,a])\ge\frac12(1-\varepsilon)\nu(B(x,r))$ or 
$\nu([a+2q'r,x+r])\ge\frac12(1-\varepsilon)\nu(B(x,r))$. 
Note that the length of 
both of these intervals is at most $r(1+\delta)$.
We assume that $\nu([a+2q'r,x+r])\ge\frac12(1-\varepsilon)\nu(B(x,r))$.
The other case can be treated similarly. Setting
$b=a+2q'r$ and $c=x+r$ \clm(essent) implies that either (1) or (2)
holds. 

Assuming the validity of (1) we find $z\in\spt(\nu)$ 
such that
$$\gamma_1\nu(B(x,r))\equiv2^{-n-1}(1-5\varepsilon)(1-\varepsilon)\nu(B(x,r))
   \le\nu(B(z,2\delta(1+\delta)r))\equiv\nu(B(z,\lambda_1r)).$$
If (2) holds instead of (1) in \clm(essent), then the assumptions of
\clm(essent) are again satisfied for the points given in (2). Assuming that
when applying \clm(essent) (2) is valid $n$ times we find
$z\in\spt(\nu)$ such that 
$$\align
  \gamma_2\nu(B(x,r))
  &\equiv\frac 12(1-2^{-n})^n(1-5\varepsilon)^n(1-\varepsilon)\nu(B(x,r))\\
  &\le\nu(B(z,(\frac{1+\delta}2)^n(1+\delta)r))\equiv\nu(B(z,\lambda_2r)).
\endalign$$
In the remaining case (2) holds $0<l<n$ times in the application of 
\clm(essent). Then there is $z\in\spt(\nu)$ with
$$\align
 \gamma_3\nu(B(x,r))
  &\equiv(1-2^{-n})^l(1-5\varepsilon)^{l+1}2^{-n-1}(1-\varepsilon)\nu(B(x,r))\\
  &\le\nu(B(z,(\frac{1+\delta}2)^l2\delta(1+\delta)r))
  \equiv\nu(B(z,\lambda_3r)).
\endalign$$

Repeating the above procedure we find $z_k\in\spt(\nu)$ for all $k\ge1$ 
and 
$(\Gamma_i,\Lambda_i)\in$
$\{(\gamma_1,\lambda_1),(\gamma_2,\lambda_2),(\gamma_3,\lambda_3)\}$
for all $1\le i\le k$ such that
$$(\prod_{i=1}^k\Gamma_i)\nu(B(x,r))\le\nu(B(z_k,(\prod_{i=1}^k\Lambda_i)r))
  \le(\prod_{i=1}^k\Lambda_i)^sr^s$$
by \equ(upnu). This gives for all $k$ 
$$s\le\frac{\sum_{i=1}^k\big(\log\Gamma_i+\frac 1k\log\nu(B(x,r))\big)}
           {\sum_{i=1}^k\big(\log \Lambda_i+\frac 1k\log r\big)}
   \equiv\frac{\sum_{i=1}^k\alpha_{i,k}}{\sum_{i=1}^k\beta_{i,k}}
   \le\max_{1\le i\le k}\frac{\alpha_{i,k}}{\beta_{i,k}}$$
(all terms are negative) implying 
$$s\le\max_{i=1,2,3}\frac{\log\gamma_i}{\log\lambda_i}.$$
The claim follows since this upper bound goes to zero as $\delta$ tends to 
zero. \qed

\medskip

\SECTION Acknowledgements

EJ and MJ thank P. Mattila for useful discussions.
This work was partly done when EJ and MJ were visiting the Department of
Mathematics at the University of Geneva. EJ acknowledges the Academy of
Finland (grant number 38288), and he and JPE the Fonds National Suisse
de la Recherche
Scientifique (grant number 20-50'493.97), and the European Science Foundation 
programme Probabilistic Methods in Non-hyperbolic Dynamics. MJ acknowledges
the financial support of the Academy of Finland and Vilho, Yrj\"o and Kalle
V\"ais\"al\"a Fund.

\bye